# Electrostatically tunable axisymmetric vibrations of soft electroactive tubes

Fangzhou Zhu[a], Bin Wu[b,*], Michel Destrade[b,a], Weiqiu Chen[a,c]

[a] Key Laboratory of Soft Machines and Smart Devices of Zhejiang Province and Department of Engineering Mechanics, Zhejiang University, Hangzhou 310027, P.R. China

[b] School of Mathematics, Statistics and Applied Mathematics, NUI Galway, University Road, Galway, Ireland

[c] Soft Matter Research Center, Zhejiang University, Hangzhou 310027, P.R. China

**ABSTRACT:** Due to their unique electromechanical coupling properties, soft electroactive (SEA) resonators are actively tunable, extremely suitable, and practically important for designing the next-generation acoustic and vibration treatment devices. In this paper, we investigate the electrostatically tunable axisymmetric vibrations of SEA tubes with different geometrical sizes. Both axisymmetric torsional and longitudinal vibrations are considered for an incompressible SEA cylindrical tube under inhomogeneous biasing fields induced by radial electric voltage and axial pre-stretch. The state-space method, which combines the state-space formalism in cylindrical coordinates with the approximate laminate technique, is employed to derive the frequency equations for two separate classes of axisymmetric vibration of the tube subjected to appropriate boundary conditions. Numerical calculations are performed to validate the convergence and accuracy of the state-space method and to illuminate that the axisymmetric vibration characteristics of SEA tubes may be tuned significantly by adjusting the electromechanical biasing fields as well as altering the tube geometry. The reported results provide a solid guidance for the proper design of tunable resonant devices composed of SEA tubes.

**Keywords**: Soft electroactive tube; axisymmetric vibrations; inhomogeneous biasing fields; state-space formalism; Tunable resonator

*Corresponding author.
E-mail address: bin.wu@nuigalway.ie (Bin Wu).

## 1. Introduction

Compared with traditional piezoelectric materials, soft electroactive (SEA) materials, besides exhibiting the exotic capability of high-speed electrical actuation with strains greater than 100% [1], also possess many other excellent electromechanical properties such as low actuation voltage, high fracture toughness and high energy density [2, 3]. These characteristics therefore have received considerable academic and industrial interest, and found widespread applications ranging from actuators, sensors and energy harvesters to biomedical and flexible electronic devices [4-8]. It is generally accepted that electric stimuli can affect the electromechanical characteristics of SEA materials in a rapid and reversible way, which in turn provides an effective approach to tune the vibration and wave characteristics of SEA structures and devices. Consequently, SEA materials can be ideally applied to the manufacturing of high-performance vibration and wave devices such as tunable resonators and acoustic/elastic waveguides [9-13].

Strong nonlinearity and electromechanical coupling of SEA materials are two important aspects in developing a general continuum mechanics framework. Early development of the nonlinear theory of electroelasticity can be tracked back to the seminal works of Toupin [14, 15] for static and dynamic analyses of finitely deformed elastic dielectrics. Later on, Tiersten [16] extended Toupin’s formulations to further incorporate thermal effects and developed a thermo-electro-elastic coupled theory by applying the laws of continuum physics to a well-defined macroscopic model. Due to the development of various smart materials and structures as well as their extensive application prospects, a general nonlinear continuum theory for electro-magneto-mechanical couplings has been regularly reformulated since 1980s [17, 18]. During this period, particular efforts have been made on finite element simulations of multifield coupling problems and micromechanics analysis of smart composites. In recent decades, the appearance of SEA materials capable of large deformations on the market [19-22] has again promoted a re-interpretation, improvement and applications of nonlinear electroelasticity [23-27]. It is noted that a nonlinear continuum framework accounting for the nonlinear interaction between mechanical and electromagnetic fields, which is well documented in the monograph by Dorfmann and Ogden [27], is appropriate for the analysis of electroactive and magnetoactive materials undergoing large deformations [28-31].

In many practical applications, the performance of intelligent systems composed of SEA materials usually depends on the biasing fields induced by, for instance, pre-stretch, internal pressure and electric stimuli. On the one hand, the biasing fields may result in instability and even failure of the systems [30, 32-34]. On the other hand, they can be exploited to actively control waves and vibrations in SEA devices. For example, experiment on the lightweight push-pull acoustic transducer consisting of dielectric elastomer (DE) films for sound generation in advanced audio systems [35] showed that the push-pull driving can suppress harmonic distortion. Hosoya et al. [36] fabricated and investigated a hemispherical breathing mode loudspeaker using a DE actuator, while Lu et al. [37] experimentally demonstrated an electrostatically tunable duct silencer using external control signals. Earlier, Dubois [38] used an electric biasing field to tune the resonance frequency of dielectric electroactive polymer (DEAP) membranes which requires no external actuators or variable elements, and observed a 77% resonant frequency reduction from the initial value. Moreover, Zhang et al. [39] put forward a vibration damper and achieved vibration attenuation by applying alternating oppositely phased voltages to a DE actuator. Consequently, tunable SEA resonators are extremely suitable for the next-generation acoustic treatment devices.

To investigate how the biasing fields influence the small-amplitude dynamic characteristics of SEA structures, different versions of linearized incremental theories [17, 18, 40-43] based on the nonlinear electroelasticity theory have been established in the literature by adopting either the Lagrangian description or the updated Lagrangian description as well as in terms of different energy density functions. By introducing three configurations to describe the general motion of an electroelastic body, Wu et al. [44] compared in detail different versions of nonlinear electroelasticity theory and associated linearized incremental theory, identified the similarities and differences between them, and concluded that these seemingly various theories are in principle equivalent without any essential difference.

Following the theory of nonlinear electroelasticity and its associated linearized incremental theory developed by Dorfmann and Ogden [40], much effort has been devoted in recent years to investigating the effects of biasing fields on small-amplitude wave propagation characteristics in SEA materials, such as bulk waves and different types of guided waves [12, 13, 28, 45-48]. More recently, the state-space method (SSM) was employed by Mao et al. [10] and Wang et al. [49] to explore the electrostatically tunable free vibration behaviors of SEA balloons and of multilayered electroactive plates, respectively. Numerical results in both papers proved that the SSM is a highly effective method for the analysis of SEA structures with inhomogeneous biasing field or multilayered configuration.

The purpose of the present study is to shed light on the effects of inhomogeneous biasing field and tube geometrical size on axisymmetric free vibrations of incompressible SEA cylindrical tubes. Both axisymmetric torsional and longitudinal vibrations (hereafter abbreviated as T vibrations and L vibrations) are considered. The biasing field is generated by applying an electric voltage difference between the two electrodes on inner and outer tube surfaces respectively, in addition to the pre-stretch in the axial direction (see Figure 1). The SSM proposed by Wu et al. [46] for the analysis of circumferential guided waves in SEA tubes is utilized here to tackle the problem of inhomogeneous biasing field.

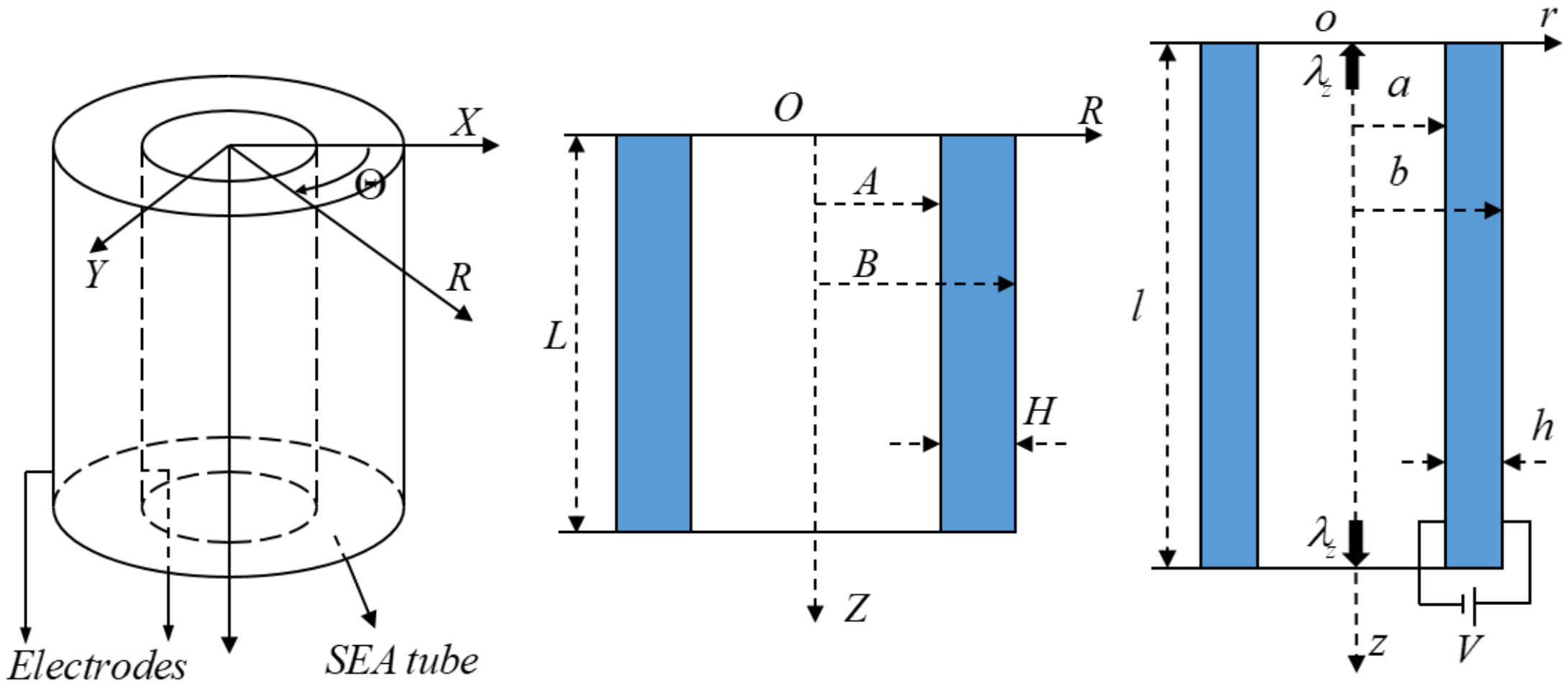

(a) (b) (c)

**Figure 1**: (a) Schematic diagram of an SEA tube with flexible surface electrodes that are mechanically negligible; (b) Undeformed configuration and related geometrical sizes; (c) Deformed configuration after activation generated by a combined action of radial electric voltage $V$ and axial pre-stretch $\lambda_z$ and related geometrical sizes.

This paper is organized as follows. Using nonlinear electroelasticity theory [23, 27],

Section 2 briefly reviews the basic formulations governing the nonlinear axisymmetric deformation and inhomogeneous biasing fields in SEA tubes characterized by a neo-Hookean ideal dielectric model. Based on the linearized incremental theory [40], Section 3 provides the governing equations and the state-space formalism in cylindrical coordinates for the incremental fields. For the generalized rigidly supported conditions, Section 4 derives the frequency equations for the two types of axisymmetric vibrations of SEA tubes with the help of the approximate laminate technique. We conduct numerical calculations in Section 5 to first, validate the convergence and accuracy of the proposed SSM for axisymmetric vibrations, and then elucidate the effects of electromechanical biasing field and tube geometry on the axisymmetric vibration characteristics. A conclusive summary is provided in Section 6 and some related mathematical expressions or derivations are presented in Appendices A-D.

## 2. Nonlinear axisymmetric deformation of an SEA tube

For better understanding the derivations of governing equations for the nonlinear axisymmetric deformation and the superimposed small-amplitude vibration in an SEA tube, the general nonlinear electroelasticity theory and its associated linearized incremental theory are briefly reviewed in Appendix A. The detailed formulations can be found in the work by Dorfmann and Ogden [23, 27, 40].

The nonlinear axisymmetric deformation of an SEA tube subjected to radial electric field, internal/external pressures and axial pre-stretch has already been provided elsewhere [45-46, 48, 50-51]. In this section, we just briefly review the basic equations and expressions when the SEA tube coated with electrodes on both the inner and outer surfaces is subjected to a radial voltage as well as an axial pre-stretch.

As displayed in Figure 1, the inner and outer radii as well as the length of the tube are specified as $A$, $B$ and $L$, respectively, in the undeformed configuration with the thickness $H = B - A$. An electric voltage difference $V$ is applied between the two surface electrodes. Meanwhile, the tube is subjected to a constant axial pre-stretch $\lambda_z$. Under these electromechanical biasing fields, the tube is deformed along with the flexible electrodes so that the inner and outer radii, the length and the thickness of the tube become $a$, $b$, $l = \lambda_z L$ and $h = b - a$, respectively.

The axisymmetric deformation for an incompressible material is given by

$$R = \sqrt{A^2 + \lambda_z \left(r^2 - a^2\right)}, \quad \theta = \Theta, \quad z = \lambda_z Z, \tag{1}$$

where $(R, \Theta, Z)$ and $(r, \theta, z)$ are cylindrical coordinates in the undeformed and deformed configurations, respectively. Thus, the deformation gradient tensor $\mathbf{F}$ can be calculated as

$$\mathbf{F} = \begin{bmatrix} \frac{\partial r}{\partial R} & \frac{\partial r}{R\partial \Theta} & \frac{\partial r}{\partial Z} \\ \frac{r\partial \Theta}{\partial R} & \frac{r\partial \theta}{R\partial \Theta} & \frac{r\partial \theta}{\partial Z} \\ \frac{\partial z}{\partial R} & \frac{\partial z}{R\partial \Theta} & \frac{\partial z}{\partial Z} \end{bmatrix} = \begin{bmatrix} \lambda_\theta^{-1}\lambda_z^{-1} & 0 & 0 \\ 0 & \lambda_\theta & 0 \\ 0 & 0 & \lambda_z \end{bmatrix}, \tag{2}$$

where $\lambda_r = \lambda_\theta^{-1}\lambda_z^{-1}$ and $\lambda_\theta = r/R$ are the radial and circumferential stretches, respectively. Accordingly, Eq. $(1)_1$ yields

$$\lambda_a^2\lambda_z - 1 = R^2\left(\lambda_\theta^2\lambda_z - 1\right)/A^2 = \left(\lambda_b^2\lambda_z - 1\right)/\eta^2, \tag{3}$$

where $\lambda_a = a/A$, $\lambda_b = b/B$ and $\eta = A/B$.

Due to the applied radial electric voltage, the biasing Eulerian electric displacement vector has only a radial component $\mathbf{D} = [D_r, 0, 0]^{\mathrm{T}}$, and thus the only non-zero component of its Lagrangian counterpart $\boldsymbol{\mathcal{D}} = \mathbf{F}^{-1}\mathbf{D}$ is $\mathcal{D}_r = \lambda_\theta\lambda_z D_r$. Furthermore, the five independent scalar invariants $I_m$ in Eq. (A.3) and the non-zero components of the total Cauchy stress tensor $\boldsymbol{\tau}$ and the Eulerian electric field vector $\mathbf{E}$ in Eq. (A.4) are

$$\begin{aligned} &I_1 = \lambda_\theta^{-2}\lambda_z^{-2} + \lambda_\theta^2 + \lambda_z^2, \quad I_2 = \lambda_\theta^2\lambda_z^2 + \lambda_\theta^{-2} + \lambda_z^{-2}, \\ &I_4 = \lambda_\theta^2\lambda_z^2 D_r^2, \quad I_5 = \lambda_\theta^{-2}\lambda_z^{-2} I_4, \quad I_6 = \lambda_\theta^{-4}\lambda_z^{-4} I_4, \end{aligned} \tag{4}$$

and

$$\begin{aligned} &\tau_{rr} = 2\lambda_\theta^{-2}\lambda_z^{-2}\left[\Omega_1 + \Omega_2\left(\lambda_\theta^2 + \lambda_z^2\right)\right] + 2\left(\Omega_5 + 2\Omega_6\lambda_\theta^{-2}\lambda_z^{-2}\right)D_r^2 - p, \\ &\tau_{\theta\theta} = 2\lambda_\theta^2\left[\Omega_1 + \Omega_2\left(\lambda_\theta^{-2}\lambda_z^{-2} + \lambda_z^2\right)\right] - p, \quad \tau_{zz} = 2\lambda_z^2\left[\Omega_1 + \Omega_2\left(\lambda_\theta^{-2}\lambda_z^{-2} + \lambda_\theta^2\right)\right] - p, \\ &E_r = 2\left(\Omega_4\lambda_\theta^2\lambda_z^2 + \Omega_5 + \Omega_6\lambda_\theta^{-2}\lambda_z^{-2}\right)D_r, \end{aligned} \tag{5}$$

where $\Omega_m = \partial\Omega/\partial I_m$, with $\Omega$ being the total energy density function.

Since the deformation is axisymmetric and also invariant along the axis, all the physical quantities are independent of the coordinates $\theta$ and $z$. As a result, Faraday's law $(A.1)_3$ is satisfied automatically, and Gauss's law $(A.1)_2$ and the equation of motion $(A.1)_1$ reduce to

$$\frac{\partial D_r}{\partial r}+\frac{D_r}{r}=\frac{1}{r}\frac{\partial\left(rD_r\right)}{\partial r}=0,\quad \frac{\partial \tau_{rr}}{\partial r}+\frac{\tau_{rr}-\tau_{\theta\theta}}{r}=0. \tag{6}$$

Integrating Eq. (6)$_1$, we obtain

$$D_r=\frac{Q(a)}{2\pi r\lambda_z L}=-\frac{Q(b)}{2\pi r\lambda_z L}, \tag{7}$$

where $Q(a)$ and $Q(b)$ are the total free surface charges on the inner and outer surfaces of the deformed SEA tube and satisfy $Q(a)+Q(b)=0$, i.e., the electrodes on the inner and outer surfaces carry equal and opposite charges. The boundary condition (A.5)$_3$ has been used to derive Eq. (7).

It is apparent from Eq. (4) that there are only three independent quantities: $\lambda_\theta$, $\lambda_z$ and $I_4$. For convenience, a reduced energy density function can be defined as

$$\Omega^*\left(\lambda_\theta,\lambda_z,I_4\right)=\Omega\left(I_1,I_2,I_4,I_5,I_6\right). \tag{8}$$

Substituting it into Eq. (5) gives

$$\lambda_\theta\Omega^*_{\lambda_\theta}=\tau_{\theta\theta}-\tau_{rr},\quad \lambda_z\Omega^*_{\lambda_z}=\tau_{zz}-\tau_{rr},\quad E_r=2\lambda_\theta^2\lambda_z^2\Omega^*_4 D_r, \tag{9}$$

where $\Omega^*_{\lambda_\theta}=\partial\Omega^*/\partial\lambda_\theta$, $\Omega^*_{\lambda_z}=\partial\Omega^*/\partial\lambda_z$ and $\Omega^*_4=\partial\Omega^*/\partial I_4$.

The electric field vector $\mathbf{E}$ is curl-free so that we can introduce an electrostatic potential $\varphi$ such that $\mathbf{E}=-\mathrm{grad}\varphi$. Then, substituting Eq. (7) into Eq. (9)$_3$ and integrating the resulting equation from the inner surface to the outer one, we obtain

$$V=\lambda_z\frac{Q(a)}{\pi L}\int_a^b \lambda_\theta^2\Omega^*_4\frac{\mathrm{d}r}{r}, \tag{10}$$

where $V=\varphi\left(a\right)-\varphi\left(b\right)$ is the electric potential difference between the inner and outer surfaces. Moreover, by inserting Eq. (9)$_1$ into Eq. (6)$_2$, conducting the integration from *a* to *b*, and assuming that both the inner and outer surfaces are traction-free, we find that

$$\int_a^b \lambda_\theta\Omega^*_{\lambda_\theta}\frac{\mathrm{d}r}{r}=0. \tag{11}$$

In a similar way, the radial normal stress can be found as

$$\tau_{rr}(r)=\int_a^r \lambda_\theta\Omega^*_{\lambda_\theta}\frac{\mathrm{d}r}{r}. \tag{12}$$

After $\tau_{rr}$ is obtained analytically or numerically from Eq. (12) for a specific energy density function, the circumferential ($\tau_{\theta\theta}$) and axial ($\tau_{zz}$) normal stresses can be derived by Eq. (9)$_1$ and (9)$_2$, respectively. Then one equation of Eq. (5)$_{1\text{-}3}$ determines the Lagrange multiplier $p$ and the resultant axial force $N$ is found by the integration of $\tau_{zz}$ over the cross-section of the deformed SEA tube.

For definiteness, a neo-Hookean ideal dielectric model is utilized to characterize the SEA tube with the (reduced) energy density functions written as

$$\begin{aligned} \Omega &= \mu(I_1-3)/2+I_5/(2\varepsilon), \\ \Omega^* &= \mu\left(\lambda_\theta^{-2}\lambda_z^{-2}+\lambda_\theta^2+\lambda_z^2-3\right)/2+\lambda_\theta^{-2}\lambda_z^{-2}I_4/(2\varepsilon), \end{aligned} \tag{13}$$

where $\mu$ denotes the shear modulus of the SEA material in the absence of biasing fields and $\varepsilon$ is the dielectric constant of the ideal dielectric material, independent of the deformation.

For the neo-Hookean ideal dielectric model, the explicit expressions of the physical variables related to the nonlinear axisymmetric deformation have been provided by Zhu et al. [50] and Wu et al. [46]. Specifically, the nonlinear axisymmetric responses governed by Eqs. (10) and (11) are

$$\begin{aligned} \bar{V} &= -\bar{Q}\frac{\eta}{1-\eta}\ln\bar{\eta}, \\ \bar{V} &= -\sqrt{\lambda_z^{-1}\left(\frac{2\lambda_a^2}{1-\bar{\eta}^2}\ln\frac{\lambda_a}{\lambda_b}+\lambda_a^2-\lambda_z^{-1}\right)}\frac{\eta}{1-\eta}\ln\bar{\eta}, \end{aligned} \tag{14}$$

where $\bar{V}=V\sqrt{\varepsilon/\mu}/H$ and $\bar{Q}=Q(a)/(2\pi A\lambda_z L\sqrt{\mu\varepsilon})$ are the dimensionless electric potential difference and surface charge, respectively, and $\bar{\eta}=a/b$ is the inner-to-outer radius ratio in the deformed configuration.

In addition, the radially inhomogeneous biasing fields required to calculate the resonant frequencies of axisymmetric vibrations are given by

$$\begin{aligned} &\lambda_\theta = \frac{\xi}{\sqrt{\eta^2/(1-\eta)^2+\lambda_z\left(\xi^2-\eta^2\lambda_a^2/(1-\eta)^2\right)}}, \quad \bar{D}_r = -\frac{\bar{V}}{\xi\ln\bar{\eta}}, \\ &\bar{p} = \lambda_z^{-1}\left[1-\ln\frac{\lambda_a}{\lambda_\theta}+\frac{\eta^2\lambda_a^2/(1-\eta)^2+\xi^2}{\left(1-\bar{\eta}^2\right)\xi^2}\ln\frac{\lambda_a}{\lambda_b}\right], \end{aligned} \tag{15}$$

where $\bar{D}_r = D_r/\sqrt{\mu\varepsilon}$ and $\bar{p}=p/\mu$ are the dimensionless radial electric displacement and Lagrange multiplier, respectively, and $\xi = r/H$ is the dimensionless radial coordinate in the deformed configuration.

## 3. Incremental equations and state-space formalism

To describe the time-dependent incremental motion accompanied by an incremental electric field in the finitely deformed SEA tube, the incremental governing equations given in Appendix A.2 are written in the cylindrical coordinates $(r,\theta,z)$ in this section. Then we reproduce the state-space formalism for the incremental fields which has been presented in Wu et al. [46].

It can be seen from Eq. (A.6)$_2$ that the incremental electric field $\dot{\mathcal{E}}_0$ is curl-free and thus an incremental electric potential $\dot{\varphi}$ can be introduced such that $\dot{\mathcal{E}}_0 = -\text{grad}\dot{\varphi}$. Its components in the cylindrical coordinates are

$$\dot{\mathcal{E}}_{0r} = -\frac{\partial\dot{\varphi}}{\partial r}, \quad \dot{\mathcal{E}}_{0\theta} = -\frac{1}{r}\frac{\partial\dot{\varphi}}{\partial\theta}, \quad \dot{\mathcal{E}}_{0z} = -\frac{\partial\dot{\varphi}}{\partial z}. \tag{16}$$

Accordingly, the incremental Gauss's law (A.6)$_3$ and the incremental equations of motion (A.6)$_1$ can be written, respectively, as

$$\frac{\partial\dot{\mathcal{D}}_{0r}}{\partial r} + \frac{1}{r}\left(\frac{\partial\dot{\mathcal{D}}_{0\theta}}{\partial\theta} + \dot{\mathcal{D}}_{0r}\right) + \frac{\partial\dot{\mathcal{D}}_{0z}}{\partial z} = 0, \tag{17}$$

and

$$\begin{aligned}
&\frac{\partial\dot{T}_{0rr}}{\partial r} + \frac{1}{r}\frac{\partial\dot{T}_{0\theta r}}{\partial\theta} + \frac{\dot{T}_{0rr} - \dot{T}_{0\theta\theta}}{r} + \frac{\partial\dot{T}_{0zr}}{\partial z} = \rho\frac{\partial^2 u_r}{\partial t^2},\\
&\frac{\partial\dot{T}_{0r\theta}}{\partial r} + \frac{1}{r}\frac{\partial\dot{T}_{0\theta\theta}}{\partial\theta} + \frac{\dot{T}_{0\theta r} + \dot{T}_{0r\theta}}{r} + \frac{\partial\dot{T}_{0z\theta}}{\partial z} = \rho\frac{\partial^2 u_\theta}{\partial t^2},\\
&\frac{\partial\dot{T}_{0rz}}{\partial r} + \frac{1}{r}\frac{\partial\dot{T}_{0\theta z}}{\partial\theta} + \frac{\partial\dot{T}_{0zz}}{\partial z} + \frac{\dot{T}_{0rz}}{r} = \rho\frac{\partial^2 u_z}{\partial t^2}.
\end{aligned} \tag{18}$$

In addition, the incremental displacement gradient tensor $\mathbf{H}$ can be written as

$$\mathbf{H} = \begin{bmatrix} \dfrac{\partial u_r}{\partial r} & \dfrac{1}{r}\left(\dfrac{\partial u_r}{\partial\theta} - u_\theta\right) & \dfrac{\partial u_r}{\partial z} \\ \dfrac{\partial u_\theta}{\partial r} & \dfrac{1}{r}\left(\dfrac{\partial u_\theta}{\partial\theta} + u_r\right) & \dfrac{\partial u_\theta}{\partial z} \\ \dfrac{\partial u_z}{\partial r} & \dfrac{1}{r}\dfrac{\partial u_z}{\partial\theta} & \dfrac{\partial u_z}{\partial z} \end{bmatrix}. \tag{19}$$

The incremental incompressibility condition (A.10) in the cylindrical coordinates thus can be expressed as

$$\frac{\partial u_r}{\partial r}+\frac{1}{r}\left(\frac{\partial u_\theta}{\partial \theta}+u_r\right)+\frac{\partial u_z}{\partial z}=0. \tag{20}$$

According to Eqs. (16) and (19), the linearized incremental constitutive equation (A.7) for incompressible SEA materials can be expressed in terms of the incremental mechanical displacement vector $\mathbf{u}$ and incremental electric potential $\dot{\varphi}$ as

$$\begin{aligned}
&\dot{T}_{0rr}=c_{11}\frac{\partial u_r}{\partial r}+c_{12}\frac{1}{r}\left(\frac{\partial u_\theta}{\partial \theta}+u_r\right)+c_{13}\frac{\partial u_z}{\partial z}+e_{11}\frac{\partial \dot{\varphi}}{\partial r}-\dot{p},\\
&\dot{T}_{0\theta\theta}=c_{12}\frac{\partial u_r}{\partial r}+c_{22}\frac{1}{r}\left(\frac{\partial u_\theta}{\partial \theta}+u_r\right)+c_{23}\frac{\partial u_z}{\partial z}+e_{12}\frac{\partial \dot{\varphi}}{\partial r}-\dot{p},\\
&\dot{T}_{0zz}=c_{13}\frac{\partial u_r}{\partial r}+c_{23}\frac{1}{r}\left(\frac{\partial u_\theta}{\partial \theta}+u_r\right)+c_{33}\frac{\partial u_z}{\partial z}+e_{13}\frac{\partial \dot{\varphi}}{\partial r}-\dot{p},\\
&\dot{T}_{0rz}=c_{58}\frac{\partial u_r}{\partial z}+c_{55}\frac{\partial u_z}{\partial r}+e_{35}\frac{\partial \dot{\varphi}}{\partial z},\quad \dot{T}_{0zr}=c_{88}\frac{\partial u_r}{\partial z}+c_{58}\frac{\partial u_z}{\partial r}+e_{35}\frac{\partial \dot{\varphi}}{\partial z},\\
&\dot{T}_{0\theta z}=c_{44}\frac{1}{r}\frac{\partial u_z}{\partial \theta}+c_{47}\frac{\partial u_\theta}{\partial z},\quad \dot{T}_{0z\theta}=c_{77}\frac{\partial u_\theta}{\partial z}+c_{47}\frac{1}{r}\frac{\partial u_z}{\partial \theta},\\
&\dot{T}_{0r\theta}=c_{66}\frac{\partial u_\theta}{\partial r}+c_{69}\frac{1}{r}\left(\frac{\partial u_r}{\partial \theta}-u_\theta\right)+e_{26}\frac{1}{r}\frac{\partial \dot{\varphi}}{\partial \theta},\\
&\dot{T}_{0\theta r}=c_{99}\frac{1}{r}\left(\frac{\partial u_r}{\partial \theta}-u_\theta\right)+c_{69}\frac{\partial u_\theta}{\partial r}+e_{26}\frac{1}{r}\frac{\partial \dot{\varphi}}{\partial \theta},
\end{aligned} \tag{21}$$

and

$$\begin{aligned}
&\dot{\mathcal{D}}_{0r}=e_{11}\frac{\partial u_r}{\partial r}+e_{12}\frac{1}{r}\left(\frac{\partial u_\theta}{\partial \theta}+u_r\right)+e_{13}\frac{\partial u_z}{\partial z}-\varepsilon_{11}\frac{\partial \dot{\varphi}}{\partial r},\\
&\dot{\mathcal{D}}_{0\theta}=e_{26}\left[\frac{1}{r}\left(\frac{\partial u_r}{\partial \theta}-u_\theta\right)+\frac{\partial u_\theta}{\partial r}\right]-\varepsilon_{22}\frac{1}{r}\frac{\partial \dot{\varphi}}{\partial \theta},\\
&\dot{\mathcal{D}}_{0z}=e_{35}\left(\frac{\partial u_z}{\partial r}+\frac{\partial u_r}{\partial z}\right)-\varepsilon_{33}\frac{\partial \dot{\varphi}}{\partial z},
\end{aligned} \tag{22}$$

where the effective material parameters $c_{ij}$, $e_{ij}$ and $\varepsilon_{ij}$ are defined as

$$\begin{aligned}
&\varepsilon_{11}=\mathcal{R}_{011}^{-1},\ \varepsilon_{22}=\mathcal{R}_{022}^{-1},\ \varepsilon_{33}=\mathcal{R}_{033}^{-1},\ e_{11}=-\Gamma_{0111}\varepsilon_{11},\ e_{12}=-\Gamma_{0221}\varepsilon_{11},\\
&e_{13}=-\Gamma_{0331}\varepsilon_{11},\ e_{26}=-\Gamma_{0122}\varepsilon_{22},\ e_{35}=-\Gamma_{0133}\varepsilon_{33},\ c_{11}=\mathcal{A}_{01111}+\Gamma_{0111}e_{11}+p,\\
&c_{12}=\mathcal{A}_{01122}+\Gamma_{0111}e_{12},\ c_{13}=\mathcal{A}_{01133}+\Gamma_{0111}e_{13},\ c_{22}=\mathcal{A}_{02222}+\Gamma_{0221}e_{12}+p,\\
&c_{23}=\mathcal{A}_{02233}+\Gamma_{0331}e_{12},\ c_{33}=\mathcal{A}_{03333}+\Gamma_{0331}e_{13}+p,\ c_{44}=\mathcal{A}_{02323},\\
&c_{47}=\mathcal{A}_{02332}+p,\ c_{55}=\mathcal{A}_{01313}+\Gamma_{0133}e_{35},\ c_{58}=\mathcal{A}_{01331}+\Gamma_{0133}e_{35}+p,\\
&c_{66}=\mathcal{A}_{01212}+\Gamma_{0122}e_{26},\ c_{77}=\mathcal{A}_{03232},\ c_{88}=\mathcal{A}_{03131}+\Gamma_{0133}e_{35},\\
&c_{69}=\mathcal{A}_{01221}+\Gamma_{0122}e_{26}+p,\ c_{99}=\mathcal{A}_{02121}+\Gamma_{0122}e_{26}.
\end{aligned} \tag{23}$$

in which the non-zero components of the instantaneous electroelastic moduli tensors $\mathcal{A}_0$,

$\boldsymbol{\Gamma}_0$ and $\mathcal{R}_0$ for the axisymmetric deformation of SEA tubes subjected to a radial electric displacement field have been derived by Wu et al. [46]. Their explicit expressions can be found in Appendix B of Ref. [46]. Note that adjusting the electromechanical biasing fields may alter the effective material properties of SEA tubes, which will generate large effects on the superimposed dynamic behavior.

It is obvious that the biasing fields are radially inhomogeneous when subjected to a radial electric voltage, which makes the effective material parameters depend on the radial coordinate $r$. Consequently, the resulting incremental governing equations are a system of coupled partial differential equations with variable coefficients, which are difficult to solve analytically or even numerically via the conventional displacement-based method. Therefore, the state-space method (SSM) [46, 52, 53] combining the state-space formalism with the approximate laminate technique is adopted in this paper to derive the frequency equations of the axisymmetric vibrations of SEA tubes.

The basic incremental governing equations (17)-(18) and (20)-(22) then can be transformed into a set of first-order ordinary differential equations as follows

$$\frac{\partial \mathbf{Y}}{\partial r} = \mathbf{M}\mathbf{Y}, \tag{24}$$

which is called the state equation, where the incremental state vector $\mathbf{Y}$ is defined as

$$\mathbf{Y} = \left[u_r, u_\theta, u_z, \dot{\varphi}, \dot{T}_{0rr}, \dot{T}_{0r\theta}, \dot{T}_{0rz}, \dot{\mathcal{D}}_{0r}\right]^{\mathrm{T}}, \tag{25}$$

and $\mathbf{M}$ is an $8\times 8$ system matrix, with its four $4\times 4$ sub-matrices presented in Appendix B.

## 4. Axisymmetric vibrations of an SEA tube

### 4.1 Approximate laminate technique

In this section, the state-space formalism is combined with the approximate laminate technique to derive the frequency equations of axisymmetric vibrations superimposed upon an activated SEA tube undergoing the finite static axisymmetric deformation described in Section 2. For the axisymmetric vibrations independent of $\theta$, the relation $\partial/\partial\theta = 0$ is fulfilled. In this case, in view of Appendix B, the state equation (24) can be simplified to

$$\frac{\partial \mathbf{Y}_k}{\partial r} = \mathbf{M}_k \mathbf{Y}_k, \quad k \in \{1, 2\}, \tag{26}$$

where $\mathbf{Y}_1 = \left[u_r, u_z, \dot{\varphi}, \dot{T}_{0rr}, \dot{T}_{0rz}, \dot{\mathcal{D}}_{0r}\right]^{\mathrm{T}}$ and $\mathbf{Y}_2 = \left[u_\theta, \dot{T}_{0r\theta}\right]^{\mathrm{T}}$ are the incremental state vectors corresponding to the axisymmetric vibrations, and

$$\mathbf{M}_1 = \begin{bmatrix} -\frac{1}{r} & -\frac{\partial}{\partial z} & 0 & 0 & 0 & 0 \\ -\frac{c_{58}}{c_{55}}\frac{\partial}{\partial z} & 0 & -\frac{e_{35}}{c_{55}}\frac{\partial}{\partial z} & 0 & \frac{1}{c_{55}} & 0 \\ \frac{q_1}{r} & q_2\frac{\partial}{\partial z} & 0 & 0 & 0 & -\frac{1}{\varepsilon_{11}} \\ \rho\frac{\partial^2}{\partial t^2}+\frac{q_3}{r^2}-q_9\frac{\partial^2}{\partial z^2} & \frac{q_4}{r}\frac{\partial}{\partial z} & -q_{10}\frac{\partial^2}{\partial z^2} & 0 & -\frac{c_{58}}{c_{55}}\frac{\partial}{\partial z} & -\frac{q_1}{r} \\ -\frac{q_5}{r}\frac{\partial}{\partial z} & \rho\frac{\partial^2}{\partial t^2}-q_6\frac{\partial^2}{\partial z^2} & 0 & -\frac{\partial}{\partial z} & -\frac{1}{r} & q_2\frac{\partial}{\partial z} \\ -q_{10}\frac{\partial^2}{\partial z^2} & 0 & q_{12}\frac{\partial^2}{\partial z^2} & 0 & -\frac{e_{35}}{c_{55}}\frac{\partial}{\partial z} & -\frac{1}{r} \end{bmatrix},$$

$$\mathbf{M}_2 = \begin{bmatrix} \frac{c_{69}}{c_{66}}\frac{1}{r} & \frac{1}{c_{66}} \\ \rho\frac{\partial^2}{\partial t^2}+\frac{q_7}{r^2}-c_{77}\frac{\partial^2}{\partial z^2} & -\left(\frac{c_{69}}{c_{66}}+1\right)\frac{1}{r} \end{bmatrix}. \tag{27}$$

It is apparent from equations (26) and (27) that the six unknown functions $u_r$, $u_z$, $\dot{\varphi}$, $\dot{T}_{0rr}$, $\dot{T}_{0rz}$ and $\dot{\mathcal{D}}_{0r}$ are uncoupled from the other two unknown functions $u_\theta$ and $\dot{T}_{0r\theta}$. Hence, there exist two independent classes of incremental axisymmetric vibrations superimposed on the underlying deformed configuration: the axisymmetric longitudinal vibrations (L vibrations) involving $\mathbf{Y}_1$ and $\mathbf{M}_1$, with the non-zero mechanical displacement components $u_r$ and $u_z$ coupled with the incremental electrical quantities (see Figures 5(b-d)); and the purely torsional vibrations (T vibrations) governed by $\mathbf{Y}_2$ and $\mathbf{M}_2$, with the sole displacement component $u_\theta$ uncoupled from the incremental electrical quantities (see Figure 6). Note that the cylindrically breathing mode characterized by the sole radial displacement $u_r$ is a special mode of the L vibrations (see Figure 5(a)), which needs to be dealt with separately.

Assume the deformed SEA tube (see Fig. 1(c)) is subject to the generalized rigidly supported (GRS) conditions [53] at the two ends. Moreover, we suppose that the electric inductions in the surrounding vacuum near the tube ends are negligible so that the zero incremental electric displacement condition applies at the tube ends. Thus, the incremental mechanical and electric boundary conditions are

$$u_z = \dot{T}_{0zr} = \dot{T}_{0z\theta} = \dot{\mathcal{D}}_{0z} = 0, \quad (z = 0, l). \tag{28}$$

For the harmonic axisymmetric free vibrations of the SEA tube, we assume that

$$\mathbf{Y}_1 = \begin{bmatrix} u_r \\ u_z \\ \dot{\varphi} \\ \dot{T}_{0rr} \\ \dot{T}_{0rz} \\ \dot{\mathcal{D}}_{0r} \end{bmatrix} = \begin{bmatrix} HU_r(\xi)\cos(n\pi\zeta) \\ HU_z(\xi)\sin(n\pi\zeta) \\ H\sqrt{\mu/\varepsilon}\Phi(\xi)\cos(n\pi\zeta) \\ \mu\Sigma_{0rr}(\xi)\cos(n\pi\zeta) \\ \mu\Sigma_{0rz}(\xi)\sin(n\pi\zeta) \\ \sqrt{\mu\varepsilon}\Delta_{0r}(\xi)\cos(n\pi\zeta) \end{bmatrix} e^{\mathrm{i}\omega t}, \quad \mathbf{Y}_2 = \begin{bmatrix} u_\theta \\ \dot{T}_{0r\theta} \end{bmatrix} = \begin{bmatrix} HU_\theta(\xi)\cos(n\pi\zeta) \\ \mu\Sigma_{0r\theta}(\xi)\cos(n\pi\zeta) \end{bmatrix} e^{\mathrm{i}\omega t}, \tag{29}$$

where $\mathrm{i} = \sqrt{-1}$ is the imaginary unit, $\omega$ is the circular frequency of vibration, $\xi = r/H$ and $\zeta = z/l$ are the dimensionless radial and axial coordinates in the deformed configuration, $\mu$ and $\varepsilon$ are material constants described in Section 2, and $n$ is the axial mode number. Note that the circumferential mode number is equal to zero for the axisymmetric vibrations. According to Eqs. $(21)_{5,7}$-$(22)_3$ and (29), the incremental boundary conditions (28) are satisfied automatically.

Substituting Eq. (29) into Eqs. (26)-(27), we obtain the dimensionless form of the state equations as

$$\frac{\mathrm{d}\bar{\mathbf{Y}}_k(\xi)}{\mathrm{d}\xi} = \bar{\mathbf{M}}_k(\xi)\bar{\mathbf{Y}}_k(\xi), \quad k \in \{1, 2\}, \tag{30}$$

where $\bar{\mathbf{Y}}_1 = [U_r, U_z, \Phi, \Sigma_{0rr}, \Sigma_{0rz}, \Delta_{0r}]^{\mathrm{T}}$ and $\bar{\mathbf{Y}}_2 = [U_\theta, \Sigma_{0r\theta}]^{\mathrm{T}}$ are the dimensionless incremental state vectors, and the dimensionless system matrices $\bar{\mathbf{M}}_k$ are written as

$$\bar{\mathbf{M}}_1 = \begin{bmatrix} -\frac{1}{\xi} & -\chi & 0 & 0 & 0 & 0 \\ \delta_1\chi & 0 & \delta_2\chi & 0 & \frac{1}{\bar{c}_{55}} & 0 \\ \frac{\bar{q}_1}{\xi} & \bar{q}_2\chi & 0 & 0 & 0 & -\frac{1}{\bar{\varepsilon}_{11}} \\ \frac{\bar{q}_3}{\xi^2} + \bar{q}_9\chi^2 - \varpi^2 & \frac{\bar{q}_4}{\xi}\chi & \bar{q}_{10}\chi^2 & 0 & -\delta_1\chi & -\frac{\bar{q}_1}{\xi} \\ \frac{\bar{q}_5}{\xi}\chi & \bar{q}_6\chi^2 - \varpi^2 & 0 & \chi & -\frac{1}{\xi} & -\bar{q}_2\chi \\ \bar{q}_{10}\chi^2 & 0 & -\bar{q}_{12}\chi^2 & 0 & -\delta_2\chi & -\frac{1}{\xi} \end{bmatrix},$$
$$\bar{\mathbf{M}}_2 = \begin{bmatrix} \frac{\delta_3}{\xi} & \frac{1}{\bar{c}_{66}} \\ \frac{\bar{q}_7}{\xi^2} + \bar{c}_{77}\chi^2 - \varpi^2 & -\frac{\delta_3 + 1}{\xi} \end{bmatrix}, \tag{31}$$

in which the dimensionless quantities are defined as follows:

$$\chi = n\pi H/l = n\pi H/(\lambda_z L), \quad \bar{c}_{ij} = c_{ij}/\mu, \quad \bar{\varepsilon}_{11} = \varepsilon_{11}/\varepsilon,$$
$$\bar{q}_j = q_j\sqrt{\varepsilon/\mu}\,(j=1,2), \quad \bar{q}_j = q_j/\mu\,(j=3-7,9), \quad \bar{q}_{10} = q_{10}/\sqrt{\mu\varepsilon}, \tag{32}$$
$$\bar{q}_{12} = q_{12}/\varepsilon, \quad \delta_1 = \bar{c}_{58}/\bar{c}_{55}, \quad \delta_2 = \bar{e}_{35}/\bar{c}_{55}, \quad \delta_3 = \bar{c}_{69}/\bar{c}_{66}, \quad \bar{e}_{35} = e_{35}/\sqrt{\mu\varepsilon},$$

and $\varpi = \omega H/\sqrt{\mu/\rho}$ is the dimensionless circular frequency. It is evident from Eqs. (30) and (31) that the dimensionless system matrices $\bar{\mathbf{M}}_k$ depend on $\xi$, making it difficult to obtain exact solutions to Eq. (30) directly. As a result, the approximate laminate technique is adopted below to obtain approximately the analytical forms of the frequency equations.

We now divide the deformed SEA tube with inner radius $a$ and thickness $h$ into $N$ equal and thin sublayers. The radial coordinates $r_{j0} = a + (j-1)h/N$ and $r_{j1} = a + jh/N$ are used to describe the inner and outer radii of the $j$-th sublayer and their corresponding dimensionless radial coordinates $\xi_{j0} = r_{j0}/H$ and $\xi_{j1} = r_{j1}/H$ are expressed as

$$\xi_{j0} = \frac{\lambda_a \eta}{1-\eta} + (j-1)\frac{\lambda_b(1-\bar{\eta})}{N(1-\eta)}, \quad \xi_{j1} = \frac{\lambda_a \eta}{1-\eta} + j\frac{\lambda_b(1-\bar{\eta})}{N(1-\eta)}. \tag{33}$$

If the number of sublayers $N$ is sufficiently large, every sublayer is thin enough so that the system matrices $\bar{\mathbf{M}}_k$ within each sublayer can be approximately regarded as constant. In the following, the values of the material parameters and the dimensionless radial coordinate itself are calculated at each mid-surface. The dimensionless radial coordinate corresponding to the mid-surface of the $j$-th sublayer is

$$\xi_{jm} = \frac{\lambda_a \eta}{1-\eta} + (2j-1)\frac{\lambda_b(1-\bar{\eta})}{2N(1-\eta)}. \tag{34}$$

Applying Eq. (30) to the $j$-th sublayer, we obtain the formal solutions as

$$\bar{\mathbf{Y}}_k(\xi) = \exp\left[(\xi - \xi_{j0})\bar{\mathbf{M}}_{kj}(\xi_{jm})\right]\bar{\mathbf{Y}}_k(\xi_{j0}),$$
$$(k=1,2; \quad \xi_{j0} \le \xi \le \xi_{j1}; \quad j = 1,2,\cdots N), \tag{35}$$

where $\bar{\mathbf{M}}_{kj}(\xi_{jm})$ are the approximated constant system matrices within the $j$-th sublayer. Setting $\xi = \xi_{j1}$ in Eq. (35), we find the following transfer relation between the incremental state vectors at the inner and outer surfaces of the $j$-th sublayer as

$$\bar{\mathbf{Y}}_k(\xi_{j1}) = \exp\left[\frac{\lambda_b(1-\bar{\eta})}{N(1-\eta)}\bar{\mathbf{M}}_{kj}(\xi_{jm})\right]\bar{\mathbf{Y}}_k(\xi_{j0}), \quad k\in\{1,2\}. \tag{36}$$

Considering the continuity conditions at each interface between the consecutive sublayers, we can finally obtain the transfer relation between the incremental state vectors $\bar{\mathbf{Y}}_k^{\text{in}}$ and $\bar{\mathbf{Y}}_k^{\text{ou}}$ at the inner and outer surfaces, as

$$\bar{\mathbf{Y}}_k^{\text{ou}} = \mathbf{P}_k \bar{\mathbf{Y}}_k^{\text{in}}, \quad k\in\{1,2\}, \tag{37}$$

where $\mathbf{P}_k = \prod_{j=N}^{1} \exp\{\lambda_b(1-\bar{\eta})\bar{\mathbf{M}}_{kj}/[N(1-\eta)]\}$ are the global transfer matrices of sixth-order ($k=1$) and second-order ($k=2$), through which the state variables at the inner and outer surfaces are connected.

### 4.2 Frequency equations

Assuming that the inner and outer surfaces of the SEA tube are traction-free and that the applied electric voltage remains unchanged during vibration, we can obtain the corresponding incremental boundary conditions (A.11) as

$$\dot{\varphi}^{\text{in}} = \dot{T}_{0rr}^{\text{in}} = \dot{T}_{0r\theta}^{\text{in}} = \dot{T}_{0rz}^{\text{in}} = \dot{\varphi}^{\text{ou}} = \dot{T}_{0rr}^{\text{ou}} = \dot{T}_{0r\theta}^{\text{ou}} = \dot{T}_{0rz}^{\text{ou}} = 0. \tag{38}$$

Substituting it into Eq. (37) yields two sets of independent linear algebraic equations as

$$\begin{bmatrix} P_{131} & P_{132} & P_{136} \\ P_{141} & P_{142} & P_{146} \\ P_{151} & P_{152} & P_{156} \end{bmatrix} \begin{bmatrix} U_r^{\text{in}} \\ U_z^{\text{in}} \\ \Delta_{0r}^{\text{in}} \end{bmatrix} = \begin{bmatrix} 0 \\ 0 \\ 0 \end{bmatrix}, \quad P_{221}U_\theta^{\text{in}} = 0, \tag{39}$$

where $P_{kij}$ are the components of the global transfer matrices $\mathbf{P}_k$. To obtain non-trivial solutions, the determinants of the coefficient matrices in Eq. (39) must be zero, i.e.,

$$\begin{vmatrix} P_{131} & P_{132} & P_{136} \\ P_{141} & P_{142} & P_{146} \\ P_{151} & P_{152} & P_{156} \end{vmatrix} = 0, \quad P_{221} = 0, \tag{40}$$

which determine the characteristic frequency equations of the two independent classes of axisymmetric vibration of the activated SEA tube for axial mode numbers $n \ge 1$.

The breathing mode ($n=0$) is a special mode which corresponds to the purely radial vibration. The state variables in Eq. $(29)_1$ are all assumed to be zero except for $u_r$, $\dot{T}_{0rr}$, $\dot{\varphi}$ and $\dot{\mathcal{D}}_{0r}$. As a result, the state equation (30) degenerates to

$$\frac{\mathrm{d}\bar{\mathbf{Y}}_3(\xi)}{\mathrm{d}\xi}=\bar{\mathbf{M}}_3(\xi)\bar{\mathbf{Y}}_3(\xi), \tag{41}$$

where $\bar{\mathbf{Y}}_3=\left[U_r,\Phi,\Sigma_{0rr},\Delta_{0r}\right]^{\mathrm{T}}$ is the dimensionless incremental state vector for the breathing mode, and the $4\times 4$ dimensionless system matrix $\bar{\mathbf{M}}_3$ can be written as

$$\bar{\mathbf{M}}_3=\begin{bmatrix} -\frac{1}{\xi} & 0 & 0 & 0 \\ \frac{\bar{q}_1}{\xi} & 0 & 0 & -\frac{1}{\bar{\varepsilon}_{11}} \\ \frac{\bar{q}_3}{\xi^2}-\varpi^2 & 0 & 0 & -\frac{\bar{q}_1}{\xi} \\ 0 & 0 & 0 & -\frac{1}{\xi} \end{bmatrix}. \tag{42}$$

Following a similar derivation to Eq. (40), we obtain the following frequency equation of the breathing mode as

$$\begin{vmatrix} P_{321} & P_{324} \\ P_{331} & P_{334} \end{vmatrix}=0, \tag{43}$$

where $P_{3ij}$ are the components of the global transfer matrix $\mathbf{P}_3$ for the breathing mode.

For the neo-Hookean ideal dielectric model (13), the explicit form of the dimensionless quantities appearing in Eqs. (31) and (32) are

$$\begin{aligned} &\bar{c}_{55}=\bar{c}_{66}=\lambda_z^{-2}\lambda_\theta^{-2},\ \bar{c}_{77}=\lambda_z^2,\ \bar{\varepsilon}_{11}=1,\ \ \delta_1=\delta_3=\lambda_z^2\lambda_\theta^2\left(\bar{p}-\bar{D}_r^2\right),\ \ \delta_2=-\bar{D}_r\lambda_z^2\lambda_\theta^2,\\ &\bar{q}_1=\bar{q}_2=2\bar{D}_r,\ \bar{q}_3=\lambda_\theta^2+2\bar{p}+\bar{D}_r^2+\lambda_z^{-2}\lambda_\theta^{-2},\ \bar{q}_4=\bar{q}_5=\bar{p}+\bar{D}_r^2+\lambda_z^{-2}\lambda_\theta^{-2},\\ &\bar{q}_6=\lambda_z^2+2\bar{p}+\bar{D}_r^2+\lambda_z^{-2}\lambda_\theta^{-2},\ \bar{q}_7=\lambda_\theta^2-\bar{D}_r^2-\lambda_z^2\lambda_\theta^2\left(\bar{D}_r^4+\bar{p}^2-2\bar{p}\bar{D}_r^2\right),\\ &\bar{q}_9=\lambda_z^2-\bar{D}_r^2-\lambda_z^2\lambda_\theta^2\left(\bar{D}_r^4+\bar{p}^2-2\bar{p}\bar{D}_r^2\right),\ \bar{q}_{10}=-\bar{D}_r\left(1-\beta_1\right),\ \ \bar{q}_{12}=\lambda_z^2\lambda_\theta^2\bar{D}_r^2+1. \end{aligned} \tag{44}$$

where $\bar{D}_r=D_r/\sqrt{\mu\varepsilon}$ and $\bar{p}=p/\mu$ are defined in Section 2.

## 5. Numerical results and discussions

Our goal is to study the axisymmetric vibration characteristics of SEA tubes, and in particular investigate how its resonant frequencies are affected by the electromechanical biasing fields (i.e., the axial pre-stretch $\lambda_z$ and the dimensionless radial electric voltage $\bar{V}$) and by the tube geometry (i.e., the inner-to-outer radius ratio $\eta$ and the length-to-thickness ratio $L/H$).

### 5.1 Nonlinear static response of the SEA tube

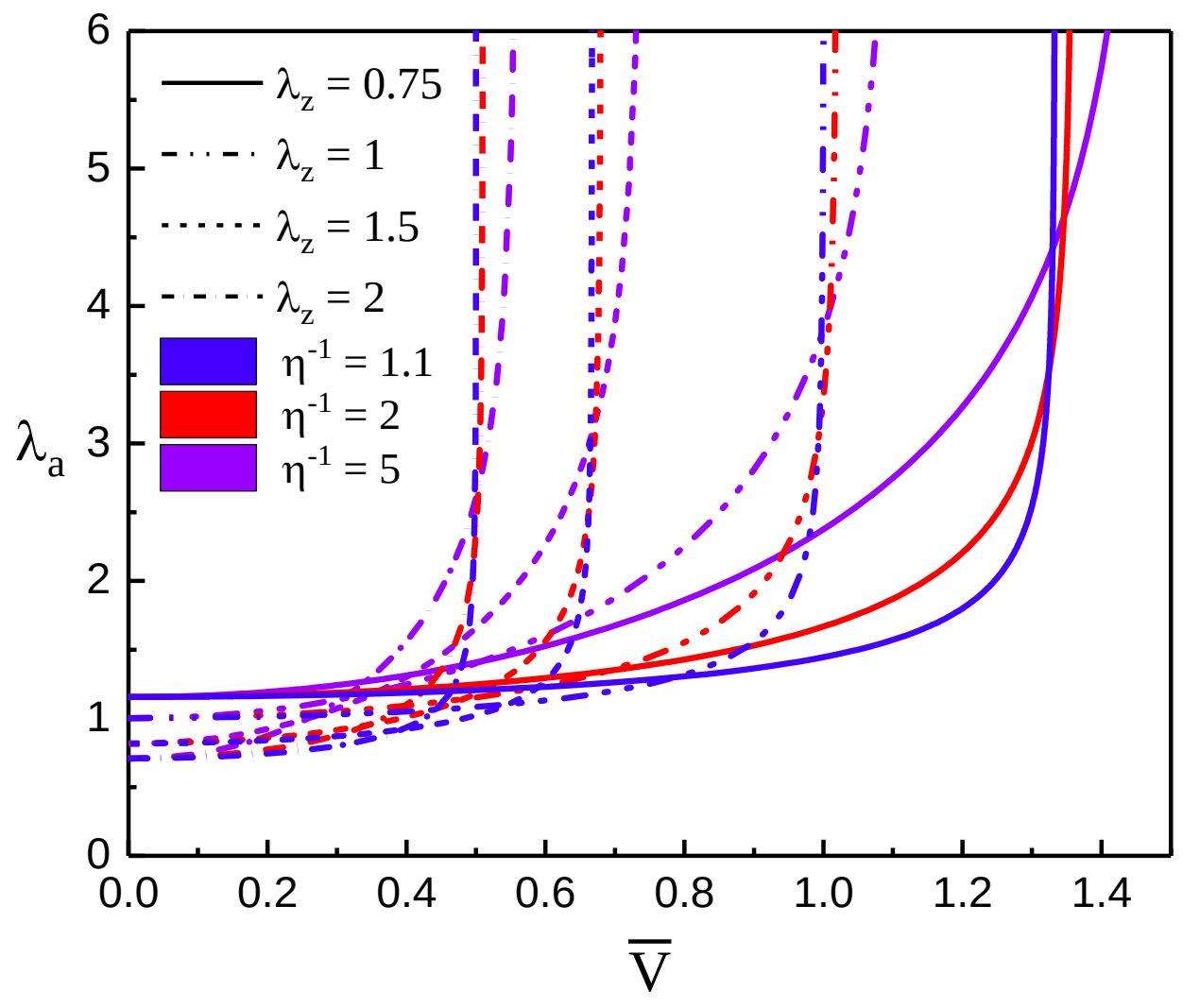

**Figure 2**: The circumferential stretch $\lambda_a$ at the outer surface versus the electric voltage $\bar{V}$ for SEA tubes with different axial pre-stretches $\lambda_z$ and outer-to-inner radius ratios $\eta^{-1}$.

Based on the nonlinear governing equation (14) for the neo-Hookean ideal dielectric model, we obtain the axisymmetric response curves in Figure 2, which displays the variations of the circumferential stretch $\lambda_a$ at the outer surface with the dimensionless voltage $\bar{V}$ for different axial pre-stretches $\lambda_z$ and inner-to-outer radius ratios $\eta$. Note that the inverse of $\eta$ (i.e., the outer-to-inner radius ratio) is chosen to be $\eta^{-1} = 1.1$, $2$ and $5$, corresponding to thin-, medium- and thick-walled tubes, respectively. The curves of $\lambda_a$ versus $\bar{V}$ for different combinations of $\lambda_z$ and $\eta^{-1}$ reveal a monotonically increasing variation trend, which means physically that the tube expands in the radial direction. We find that no axisymmetric solution exists and that the SEA tube collapses when the voltage exceeds a critical value, which we call the electromechanical instability voltage $\bar{V}_{\mathrm{EMI}}$. There, the balance between the compressive force caused by the radial electric voltage and the mechanical resistance force cannot be maintained [45, 46]. Moreover, when the voltage reaches $\bar{V}_{\mathrm{EMI}}$, a rapid rise of the curves is observed. For a fixed $\eta^{-1}$, the axial compression results in a higher $\bar{V}_{\mathrm{EMI}}$, while for the thinner SEA tube, it is easier to arrive at the electromechanical instability with a given $\lambda_z$. Specifically, the critical voltage values $\bar{V}_{\mathrm{EMI}}$ for different combinations of $\lambda_z$ and $\eta^{-1}$ are exhibited in Table 1.

**Table 1**: Electromechanical instability voltages $\bar{V}_{\mathrm{EMI}}$ for different combinations of axial pre-stretches $\lambda_z$ and outer-to-inner radius ratios $\eta^{-1}$.

| $\lambda_z$ | 0.75 | 1 | 1.5 | 2 |
|---|---|---|---|---|
| $\eta^{-1}=1.1$ | 1.333 | 1.000 | 0.666 | 0.500 |
| $\eta^{-1}=2$ | 1.358 | 1.019 | 0.679 | 0.509 |
| $\eta^{-1}=5$ | 1.467 | 1.100 | 0.734 | 0.551 |

### 5.2 Validation of the state-space method

As stated in Sections 3 and 4, the state-space method (SSM) combining the state-space formalism with the approximate laminate technique is an analytical but approximate method. It is necessary to validate its convergence and accuracy for the axisymmetric vibrations of SEA tubes.

For the convergence analysis, Tables 2-5 exhibit the variations with the number of discretized layers (NOL) of the first two dimensionless resonant frequencies of two classes of vibration for the axial mode number $n=1$ calculated by the SSM. The results for the thick and short tube are displayed in Tables 2 and 4, while Tables 3 and 5 correspond to the results for the thin and slender tubes. Obviously, the results based on the SSM show an excellent convergence rate with increasing layer number, and thus we are satisfied that we can obtain accurate resonant frequencies with an arbitrary precision via the present SSM.

**Table 2**: The first two resonant frequencies $\varpi$ of L vibration with $n=1$ of the *thick and short* tube ($\eta^{-1}=5$ and $L/H=2.5$) based on the SSM with different numbers of discretized layers (NOL) ($\lambda_z=2$ and $\bar{V}=0.3$).

| NOL | 20 | 40 | 60 | 80 | 100 | 120 | 140 | 160 |
|---|---|---|---|---|---|---|---|---|
| 1st | 1.38275 | 1.38271 | 1.38271 | 1.3827 | 1.3827 | 1.3827 | 1.3827 | 1.3827 |
| 2nd | 3.10784 | 3.11411 | 3.1153 | 3.11572 | 3.11591 | 3.11602 | 3.11608 | 3.11613 |

**Table 3**: The first two resonant frequencies $\varpi$ of L vibration with $n=1$ of the *thin and slender* tube ($\eta^{-1}=1.1$ and $L/H=100$) based on the SSM with different numbers of discretized layers (NOL) ($\lambda_z=2$ and $\bar{V}=0.3$).

| NOL | 20 | 40 | 60 | 80 | 100 | 120 | 140 | 160 |
|---|---|---|---|---|---|---|---|---|
| 1st | 0.034475 | 0.034475 | 0.034475 | 0.034475 | 0.034475 | 0.034475 | 0.034475 | 0.034475 |
| 2nd | 0.155983 | 0.155983 | 0.155983 | 0.155983 | 0.155983 | 0.155983 | 0.155983 | 0.155983 |

**Table 4**: The first two resonant frequencies $\varpi$ of T vibration with $n=1$ of the *thick and short* tube ($\eta^{-1}=5$ and $L/H=2.5$) based on the SSM with different numbers of discretized layers (NOL) ($\lambda_z=2$ and $\bar{V}=0.3$).

| NOL | 20 | 40 | 60 | 80 | 100 | 120 | 140 | 160 |
|---|---|---|---|---|---|---|---|---|
| 1st | 1.2675 | 1.26765 | 1.26768 | 1.26769 | 1.26769 | 1.2677 | 1.2677 | 1.2677 |
| 2nd | 4.10631 | 4.10741 | 4.10762 | 4.10769 | 4.10772 | 4.10774 | 4.10775 | 4.10776 |

**Table 5**: The first two resonant frequencies $\varpi$ of T vibration with $n=1$ of the *thin and slender* tube ($\eta^{-1}=1.1$ and $L/H=100$) based on the SSM with different numbers of discretized layers (NOL) ($\lambda_z=2$ and $\bar{V}=0.3$).

| NOL | 20 | 40 | 60 | 80 | 100 | 120 | 140 | 160 |
|---|---|---|---|---|---|---|---|---|
| 1st | 0.034475 | 0.034475 | 0.034475 | 0.034475 | 0.034475 | 0.034475 | 0.034475 | 0.034475 |
| 2nd | 3.14552 | 3.14552 | 3.14552 | 3.14552 | 3.14552 | 3.14552 | 3.14552 | 3.14552 |

When there is no electric voltage applied to the tube, the deformation of the pre-stretched SEA tube is homogeneous, i.e., the biasing fields become homogeneous without the application of voltage. Therefore, the exact resonant frequencies for the superimposed non-axisymmetric vibrations (including the axisymmetric vibration as a special case) can be obtained through the conventional displacement method; these frequencies are provided in Appendix C in detail. Based on the exact solutions and the SSM, the curves of the first five dimensionless resonant frequencies $\varpi$ versus the axial mode number $n$ are plotted in Figures 3(a) and 3(b) for the L and T vibrations, respectively, in the pre-stretched SEA thick and short tube ($\lambda_z=2$, $\eta^{-1}=5$ and $L/H=2.5$). The lines in Figure 3 correspond to the exact solutions while the symbols represent the solutions from the SSM. The tube is divided into 120 thin sublayers to ensure a balance between accuracy and computational speed. It is clear that resonant frequencies obtained by the SSM agree very well with the exact solutions in the entire axial mode number range for the axisymmetric vibrations. This in turn confirms the accuracy of the SSM. It should be emphasized that due to the incompressibility of the tube, there exists only one radial vibration frequency for the breathing mode $n=0$. The frequency equation of the breathing mode for an arbitrary energy function is provided in Eq. (C.20). In particular, Eq. (C.21) gives the resonant frequency of the breathing mode for the pre-stretched neo-Hookean tube. Moreover, it can be seen from Eq. (C.16) that the axial pre-stretch has no effect on the resonant frequencies of the T vibration for the neo-Hookean hyperelastic model; they are only determined by the outer-to-inner radius ratio and the length-to-thickness ratio.

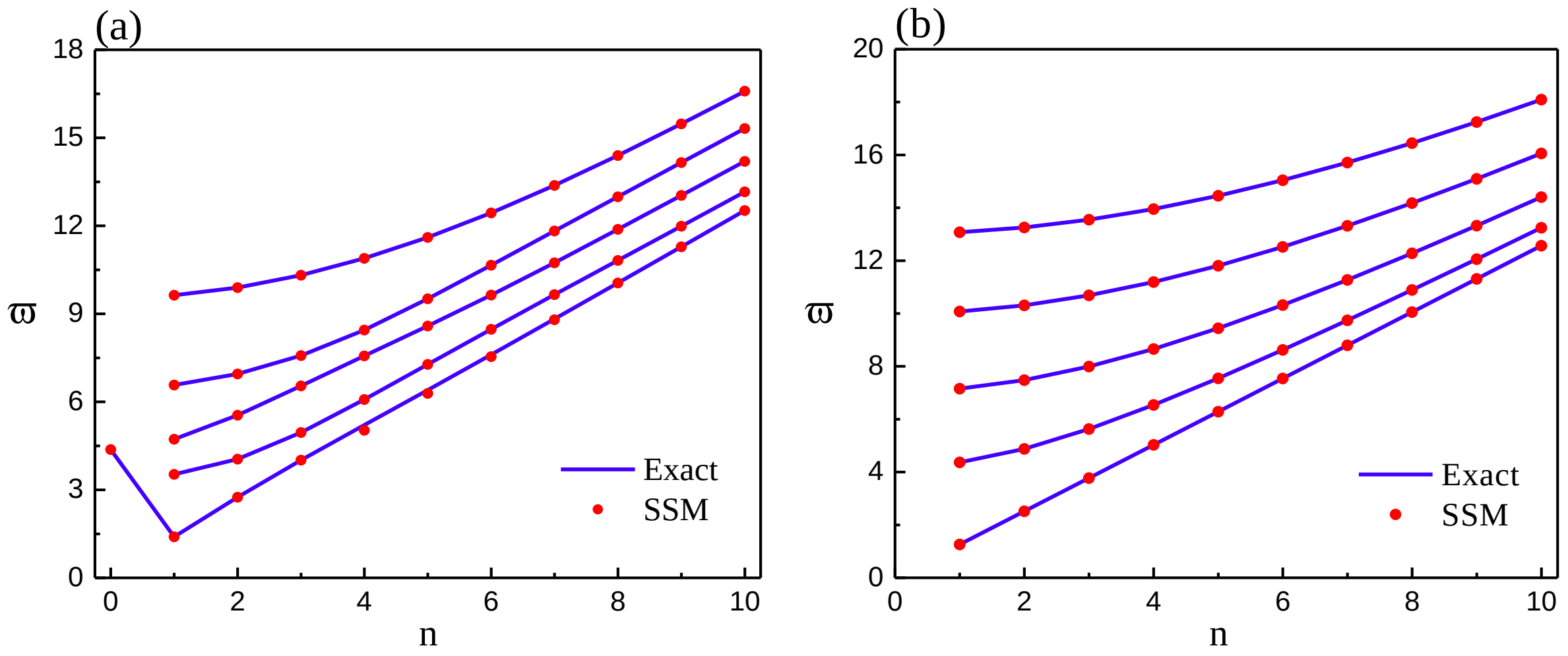

**Figure 3**: Accuracy analysis of the first five dimensionless vibration frequencies $\varpi$ of the L vibration (a) and T vibration (b) obtained by the exact solutions and the SSM for the pre-stretched ($\lambda_z = 2$) SEA *thick and short* tube ($\eta^{-1} = 5$ and $L/H = 2.5$) without voltage.

In summary, the superior convergence rate and the excellent agreement with the exact solutions demonstrate that the obtained numerical results based on the SSM are highly accurate.

### 5.3 Effect of the electromechanical biasing fields

In this subsection, we focus on how the electromechanical biasing fields influence the resonant frequencies of the two classes of axisymmetric vibration of SEA tubes. Without loss of generality, the tube geometry is taken as $\eta^{-1} = 5$ and $L/H = 2.5$ corresponding to a *thick and short* SEA tube. The number of discretized layers is set to 120 as explained in Section 5.2.

Variations of the dimensionless resonant frequency $\varpi$ with the axial mode number $n$ are displayed in Figures 4(a) and 4(b) for the L and T vibrations, respectively, for axial pre-stretch $\lambda_z = 2$ and different voltages below the electromechanical instability voltage $\bar{V}_{\mathrm{EMI}}$. It can be seen from Figure 4(a) that the first resonant frequency of L vibrations decreases at first and then grows monotonously with increasing axial mode number $n$. The lowest vibration frequency is taken at the axial mode number $n = 1$. In addition, a higher electric voltage leads to a lower vibration frequency. Specifically, when the dimensionless radial electric voltage is less than 0.2, the applied electric voltage barely affects the L vibration frequency. However, a high electric voltage surpassing 0.2 will significantly decrease the resonant frequency due to the rapid expansion of the tube as shown in Figure 2. When the applied electric voltage $\bar{V} = 0.49$ gradually approaches $\bar{V}_{\text{EMI}} = 0.55$, the effect of electromechanical coupling on the L vibration frequency is the greatest.

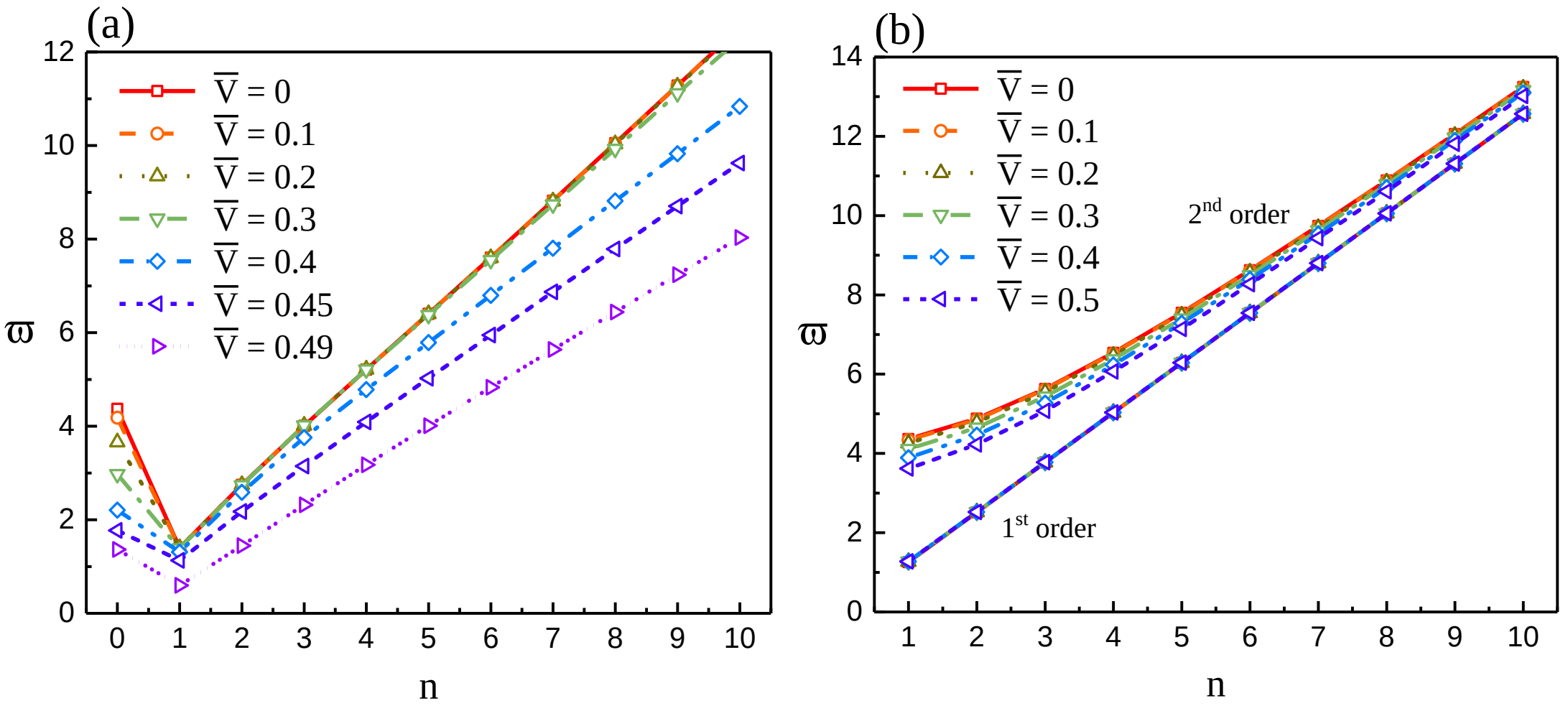

**Figure 4**: Dimensionless frequency spectra ($\varpi$ versus $n$) in a pre-streched ($\lambda_z = 2$) SEA *thick and short* tube ($\eta^{-1} = 5$ and $L/H = 2.5$) for different values of radial electric voltage: (a) the first-order frequency of L vibrations; (b) the first two frequencies of T vibrations.

For the T vibrations depicted in Figure 4(b), the change of electric voltage makes no difference to the first-order vibration frequency. This is because the first-order resonant frequency of the T vibration is governed by the relation $\rho\omega^2 - \gamma^2 c_{77} = 0$ with $\gamma = n\pi/l$ for an arbitrary energy function, which is independent of the applied voltage as verified in Appendix D. Furthermore, the first-order vibration frequency grows monotonously and linearly with increasing axial mode number. However, the second-order vibration frequency declines with the applied voltage and grows nonlinearly with the increasing axial mode number, as shown in Figure 4(b).

The first-order L vibration modes corresponding to different axial mode numbers $n = 0$, 1, 2 and 6 are presented in Figure 5. The breathing mode in Figure 5(a) has the sole radial displacement component $u_r$ that is independent of $z$ and $\theta$. According to Eqs. (C.17) and (C.19), the radial displacement for the breathing mode is an inversely proportional function of the radius. Therefore, the displacement at the inner surface is larger than that at the outer surface and the circumferential gridlines become sparse from the inner surface to the outer surface. Figures 5(b-d) show the L vibration modes for non-zero axial mode numbers ($n \neq 0$) with both the radial and axial displacement components $u_r$ and $u_z$. The SEA tube for these modes vibrates in the form of trigonometric function in the axial direction, which conforms well with the formal solutions $(29)_1$. Obviously, the axial mode number $n \neq 0$ is equal to the integer multiple of the half-wave number.

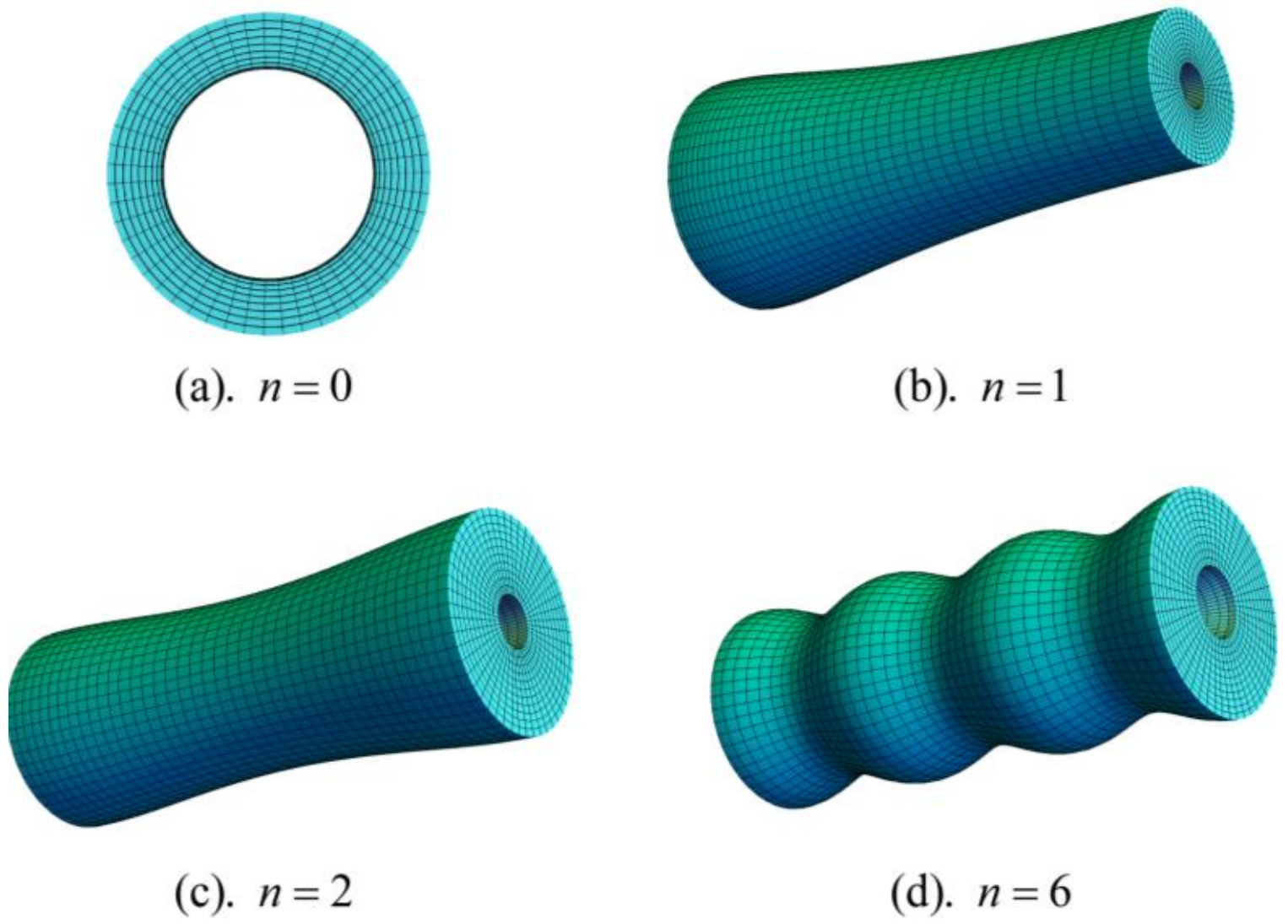

**Figure** 5: The first-order mode of the L vibrations in a pre-streched ($\lambda_z = 2$) SEA *thick and short* tube ($\eta^{-1} = 5$ and $L/H = 2.5$) with $\bar{V} = 0.2$: (a) breathing mode ($n = 0$); (b) $n = 1$; (c) $n = 2$; (d) $n = 6$.

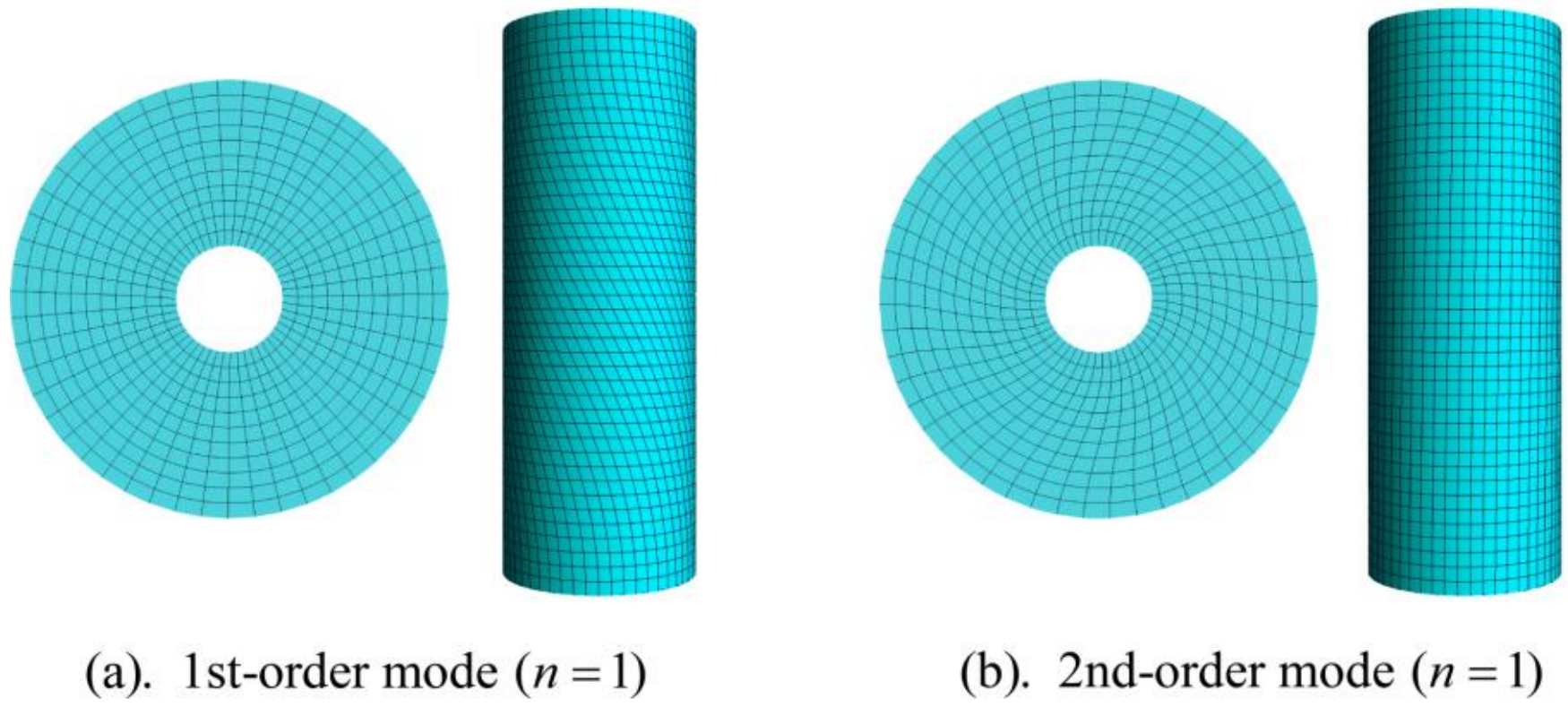

**Figure** 6: The first two modes of the T vibration with $n = 1$ in a pre-stretched ($\lambda_z = 2$) SEA *thick and short* tube ($\eta^{-1} = 5$ and $L/H = 2.5$) with $\bar{V} = 0.2$: (a) the first-order mode; (b) the second-order mode.

Figure 6 exhibits the first two T vibration modes for a fixed axial mode number $n = 1$ in an SEA tube subjected to biasing fields. For the first-order vibration mode in Figure 6(a), the sole torsional displacement component $u_\theta$ is proportional to the radius (see Eqs. (D.1) and (D.9)), and the vibration is a rotation of each cross-section as a whole about its center, which is analogous to the torsional waves in an isotropic elastic cylinder [54]. Thus, the gridlines still distribute uniformly in the cross-section. But the torsional displacement varies according to the trigonometric function in the axial direction (see the right view of Figure 6(a)). For the second-order vibration mode in Figure 6(b), the torsional displacement presents the nonlinear distribution and there exists one zero-crossing point in the mode profile along the radial direction.

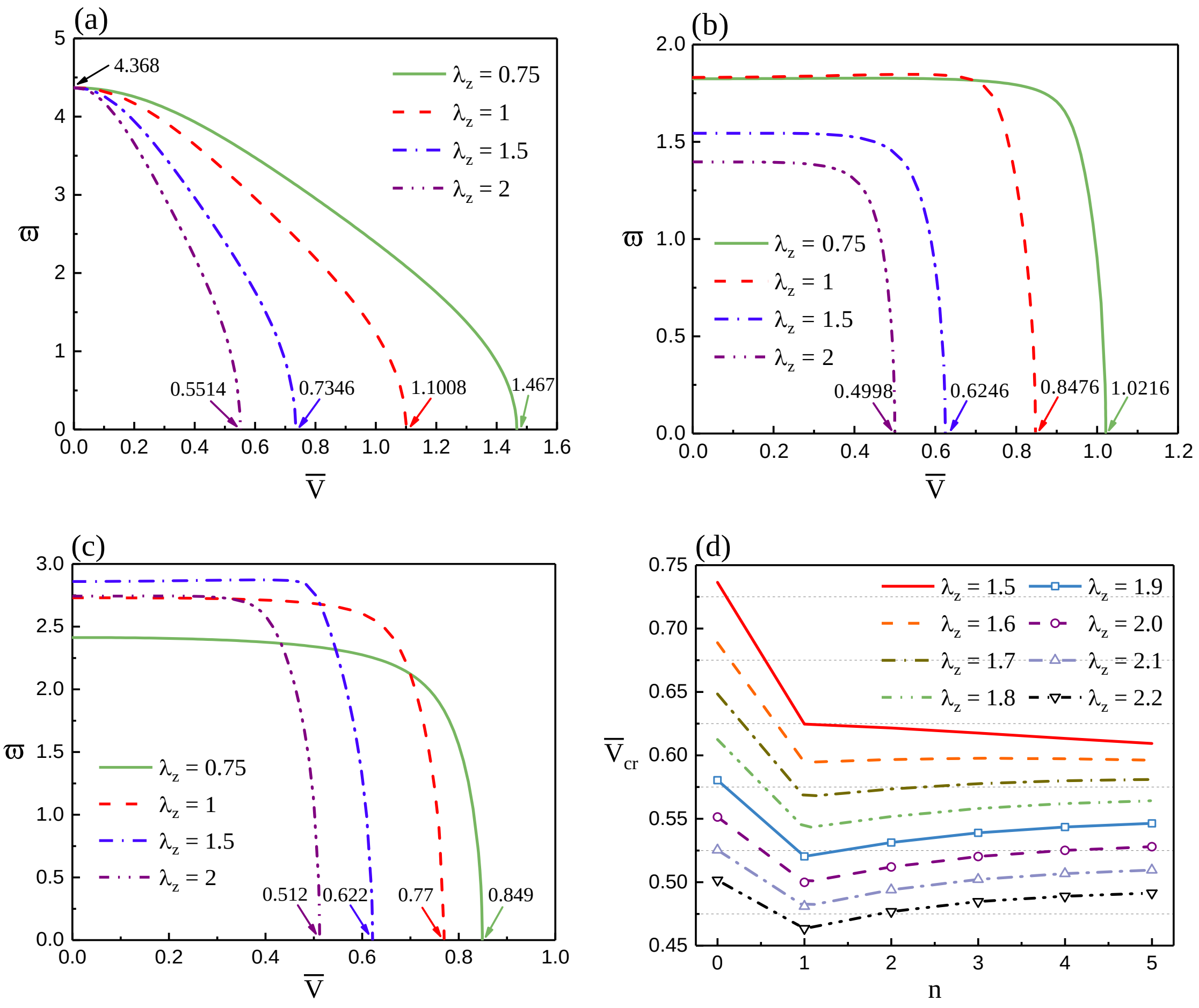

**Figure 7**: The first-order resonant frequencies $\varpi$ of the L vibration as functions of the radial electric voltage $\bar{V}$ for a *thick and short* SEA tube ($\eta^{-1} = 5$ and $L/H = 2.5$) under different axial pre-stretches $\lambda_z$: (a) breathing mode ($n = 0$); (b) $n = 1$; (c) $n = 2$. (d) Variation trend of the critical voltage $\bar{V}_{cr}$ corresponding to the instabilities with the axial mode number $n$ for different axial pre-stretches.

In order to clearly demonstrate the effect of voltage on the vibration behavior, the curves of the L vibration frequency as functions of the voltage are displayed in Figures 7(a-c) for axial mode number $n=0$, $n=1$ and $n=2$, respectively, under different axial pre-stretches. It is apparent that the resonant frequencies $\varpi$ for all three modes monotonically decrease to zero with the increasing voltage from zero to the critical voltage $\bar{V}_{cr}$ of the corresponding vibration mode. At the critical voltage $\bar{V}_{cr}$, the point $\varpi=0$ represents the axisymmetric instability of the corresponding vibration mode of the SEA tube. The decrease of the resonant frequencies is mainly due to the global stiffness of the SEA tube gradually decreasing with increasing voltage [10]. In particular, when the voltage approaches the critical value $\bar{V}_{cr}$, the global stiffness reduces rapidly so that barrelling instabilities [52, 55] occur in the SEA tube. This is why the frequencies of modes $n=1$ and $n=2$ in Figures 7(b-c) decrease gently at first and then dramatically. Moreover, we find that the larger axial pre-stretch the tube is subjected to, the lower critical voltage the tube may withstand.

For the breathing mode $n=0$ in Figure 7(a), when there is no radial electric voltage ($\bar{V}=0$), the vibration frequencies are identical for any axial pre-stretch $\lambda_z$ in the neo-Hookean SEA tube according to Eq. (C.21), and they depend only on the inner-to-outer radius ratio $\eta$. Additionally, the critical voltages corresponding to the breathing mode for different axial pre-stretches are identical to the electromechanical instability voltage $\bar{V}_{\mathrm{EMI}}$ shown in Table 1 for the axisymmetric deformation.

Furthermore, we see from Figures 7(a-c) that the critical voltage decreases monotonously with increasing axial mode number for $\lambda_z=0.75$, 1 and 1.5, while the critical voltage presents a different variation trend for $\lambda_z=2$. Therefore, Figure 7(d) demonstates the variation trend of the critical voltage $\bar{V}_{cr}$ (where the barrelling instabilities happen) with the axial mode number $n$ for different axial pre-streches from 1.5 to 2.2. With the assistance of the dotted grey auxiliary line in Figure 7(d), we see that the critical voltage decreases to a minimum at $n=1$ and then increases monotonically when the applied axial pre-stretch is over approximately 1.6. Therefore, for a higher axial pre-stretch, the SEA tube undergoes the barrelling instability first at $n=1$. If the axial pre-stretch is less than 1.6, then the SEA tube will have the barrelling instability at a higher axial mode number.

For the T vibration with $n=1$, Figure 8 displays the variations of the first two resonant frequencies with the applied voltage for different axial pre-stretches. Compared with the curves of L vibrations in Figure 7, the frequency variation trend for the T vibration in Figure 8 is quite unique. Specifically, the first-order resonant frequency (the lower curves shown in Figure 8) is independent of the applied voltage $\bar{V}$ and axial pre-stretch $\lambda_z$, as explained in Appendix D. The torsional mode of the first-order frequency exhibits a linear displacement distribution along the radial direction. Note that different $\lambda_z$ result in different electromechanical instability voltage $\bar{V}_{\mathrm{EMI}}$, as shown in Table 1 and Figure 8. The independence of the first-order vibration frequency on the biasing fields could be exploited to design a torsional resonator with a consistent working performance. However, the second-order frequency (the higher curves depicted in Figure 8) decreases gradually with the voltage, which is in accordance with the phenomenon shown in Figure 4(b).

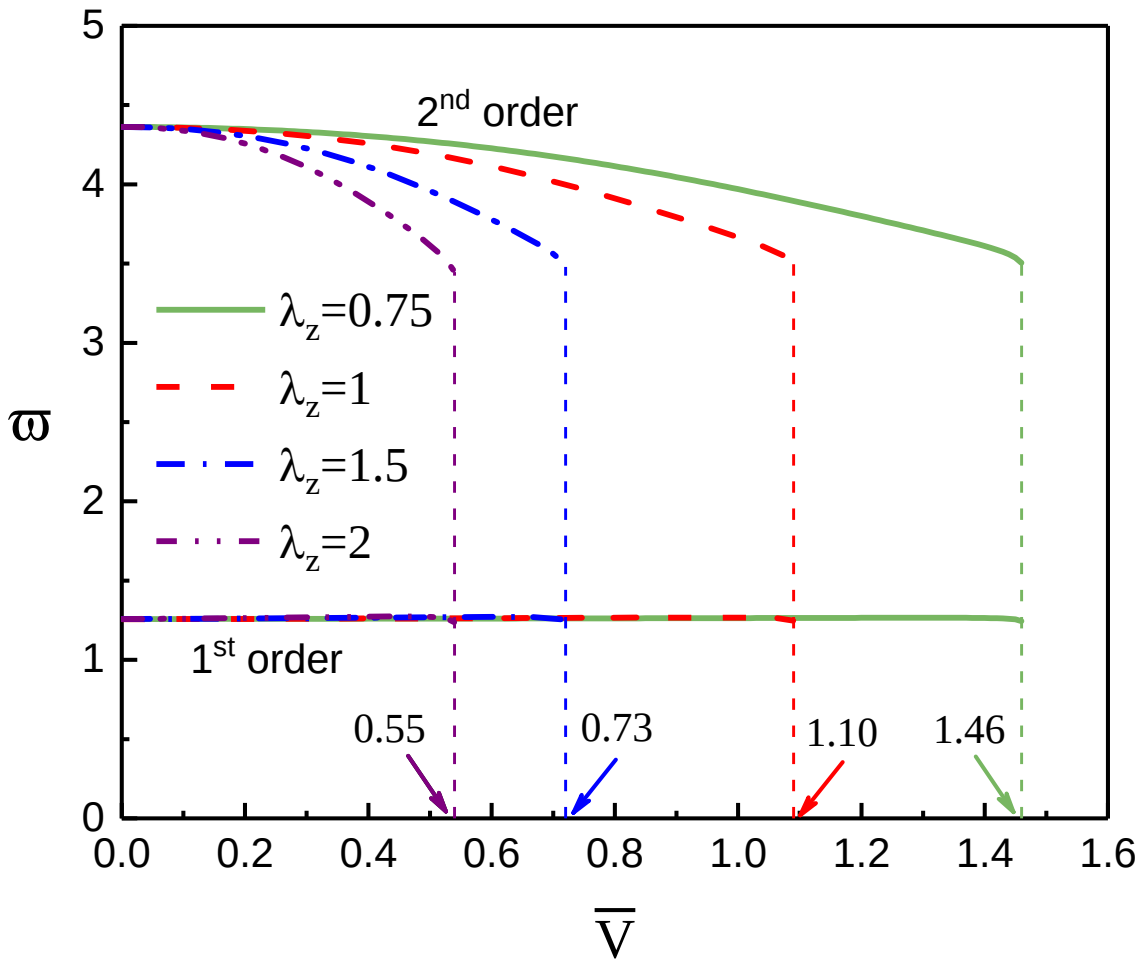

**Figure 8**: The first two resonant frequencies $\varpi$ of the T vibration with $n=1$ as functions of the radial electric voltage $\bar{V}$ for a *thick and short* SEA tube ($\eta^{-1}=5$ and $L/H=2.5$) under different axial pre-stretches $\lambda_z$.

In short, the dependence of the resonant frequency of the axisymmetric vibrations on the electromechanical biasing fields provides a possibility to tune the small-amplitude free vibrations of SEA tubes.

### 5.4 Effect of the tube geometry

Now we turn to explore how the tube geometry including the length-to-thickness ratio $L/H$ and the outer-to-inner radius ratio $\eta^{-1}$ influences the axisymmetric vibrations of the SEA tube. In this subsection, the dimensionless electric voltage is always taken as $\bar{V}=0.2$ and the tube is divided into 120 thin sublayers to guarantee convergence and accuracy.

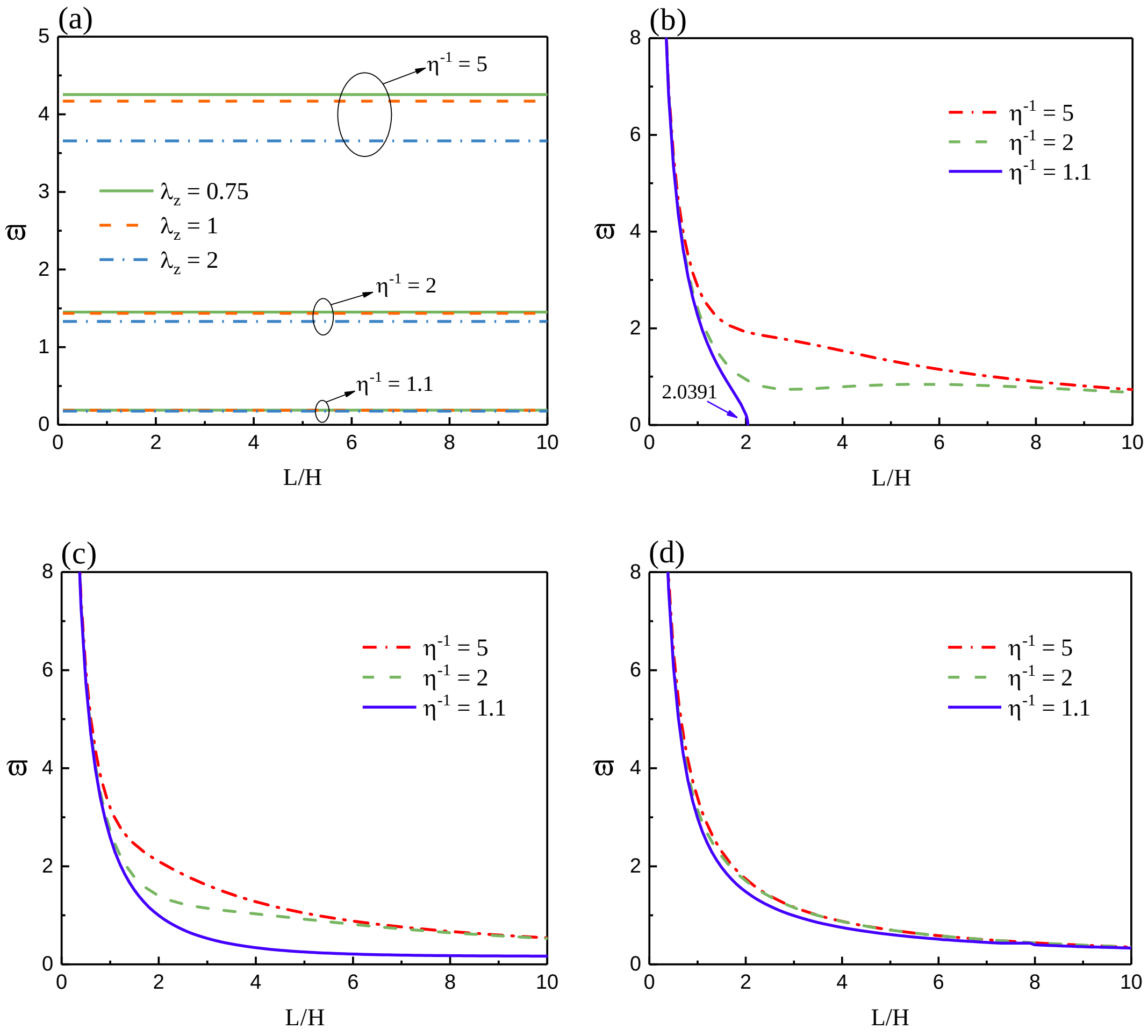

**Figure 9**: The first-order resonant frequency $\varpi$ of the L vibration as functions of the length-to-thickness ratio $L/H$ for different combinations of $\eta^{-1}$ and $\lambda_z$ with $\bar{V}=0.2$: (a) breathing mode $(n=0)$; (b) $n=1$ for $\lambda_z=0.75$; (c) $n=1$ for $\lambda_z=1$; (b) $n=1$ for $\lambda_z=2$.

For the breathing mode ($n=0$), the curves of the first-order dimensionless frequency $\varpi$ versus $L/H$ are displayed in Figure 9(a) for different combinations of $\eta^{-1}$ and $\lambda_z$. Apparently, the length-to-thickness ratio makes no difference to the vibration frequency; this is because $L/H$ disappears from the frequency equation for the breathing mode $n=0$ according to Eq. (32)$_1$. Nonetheless, the outer-to-inner radius ratio $\eta^{-1}$ and the axial pre-stretch $\lambda_z$ still have an influence on the vibration frequency for a non-zero voltage. Specifically, for a fixed axial pre-stretch, the thicker the SEA tube is, the larger resonant frequency we obtain. Additionally, increasing the axial pre-stretch lowers the resonant frequency. However, the variation gap between the axial pre-stretch and the vibration frequency depends on $\eta^{-1}$. The vibration frequency is barely affected by the axial pre-stretch for a thin tube ($\eta^{-1}=1.1$), but it is remarkably decreased by the axial pre-stretch for a thick tube ($\eta^{-1}=5$).

For the L vibration mode $n=1$, Figures 9(b)-(d) demonstrate the variations of the lowest resonant frequency with $L/H$ for different combinations of $\eta^{-1}$ and $\lambda_z$. Generally, the vibration frequency gradually decreases with increasing $L/H$ and the vibration frequency of the thick tube is higher than that of the thin tube. In addition, we see from Figures 9(b)-(d) that the curves of these three tubes with different $\eta^{-1}$ get closer to each other when increasing the axial pre-stretch. That is to say, a larger axial pre-stretch weakens the effect of the tube geometry on the L vibration behavior. It is interesting to note from Figure 9(b) that, for an axial compression $\lambda_z=0.75$, the resonant frequency of the thin tube reduces to zero when $L/H$ approaches a critical value 2.0391, which corresponds to the barrelling instability with the axial mode number $n=1$. Thus, the axisymmetric instability occurs more easily for a slender tube subjected to the axial compression.

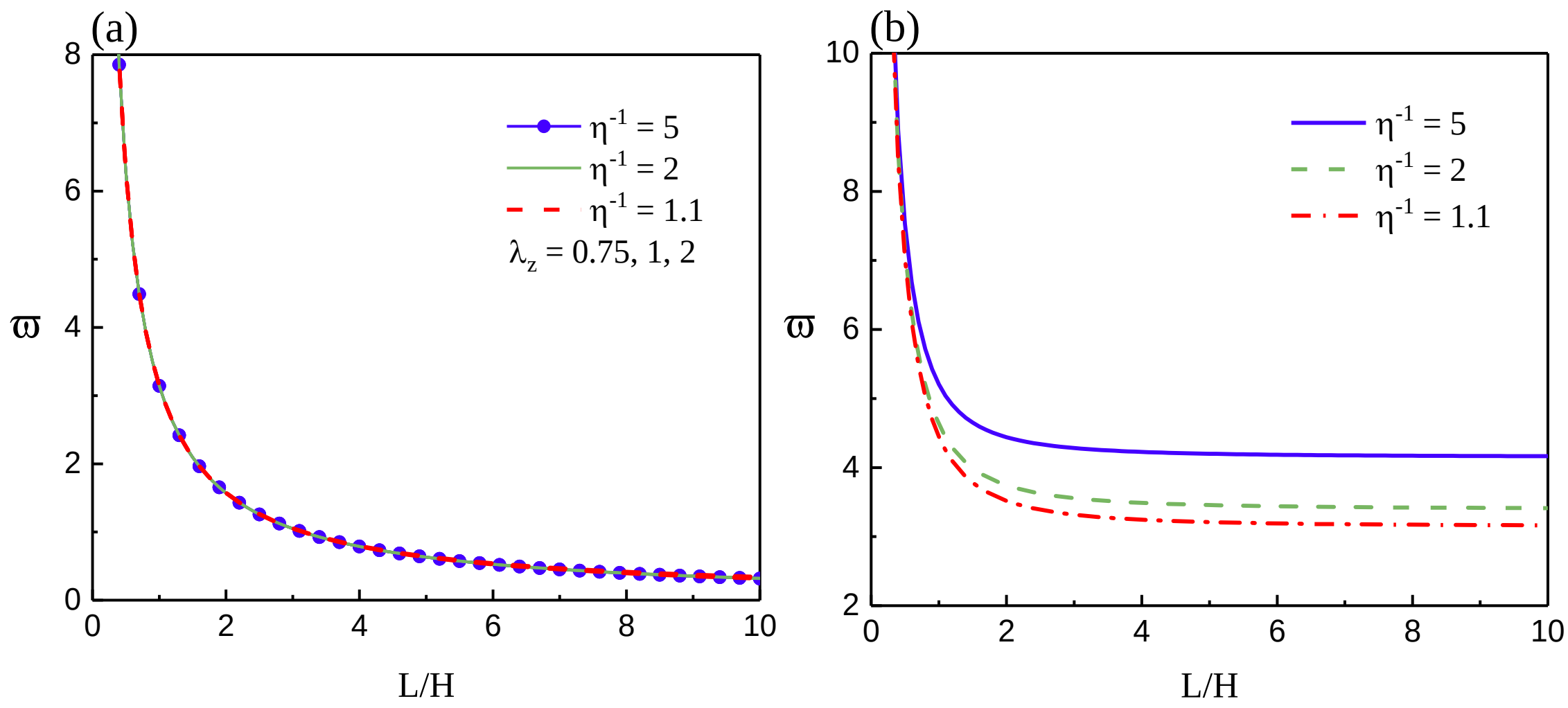

**Figure 10**: The first two resonant frequencies $\varpi$ of the T vibration with $n=1$ as functions of the length-to-thickness ratio $L/H$ for different combinations of $\eta^{-1}$ and $\lambda_z$ with $\bar{V}=0.2$: (a) the first-order frequency; (b) the second-order frequency for $\lambda_z=1$.

When it comes to the T vibration with $n=1$, Figure 10 displays the first two resonant frequencies versus $L/H$ for different combinations of $\eta^{-1}$ and $\lambda_z$. Similar to the results of L vibrations, the vibration frequency decreases monotonically with increasing $L/H$. For the first-order torsional mode shown in Figure 10(a), the resonant frequency for the neo-Hookean SEA tube satisfies $\varpi = \pi / (L / H)$ (see Eq. (D.10) in Appendix D) and is an inversely proportional function of the length-to-thickness ratio $L / H$. Therefore, the frequency is independent of the axial pre-stretch and the outer-to-inner radius ratio, as shown in Figure 10(a). According to Appendix D, the torsional displacement is distributed linearly along the radial direction. In addition, the resonant frequency tends to zero for an infinite SEA tube, which means physically that the longer tube achieves torsional instability more easily. For the second-order frequency depicted in Figure 10(b), a thicker SEA tube ($\eta^{-1} = 5$) results in a higher resonant frequency, especially when $L / H \geq 1$. Besides, the frequency is hardly affected by the axial pre-stretch for a low voltage ($\bar{V} = 0.2$), but for a higher voltage, increasing the axial pre-stretch will lead to a lower resonant frequency, which is analogous to the phenomena observed in Figure 8.

## 6. Conclusions

In this work, we conducted an analytical study of the small-amplitude axisymmetric vibrations of SEA tubes subjected to inhomogeneous biasing fields. The theory of nonlinear electroelasticity and the associated linearized incremental theory constitute the basis of our analysis. To tackle the problem of inhomogeneity in the deformed configuration, we adopted the state-space method (SSM) which combines the state-space formalism in cylindrical coordinates with the approximate laminate technique. We obtained the characteristic frequency equations for two independent classes of axisymmetric vibrations (i.e., L and T vibrations) by imposing proper mechanical and electric boundary conditions. To confirm the accuracy of the SSM, we also employed the displacement method to derive the exact frequency equations of the axisymmetric vibrations in a pre-stretched hyperelastic tube. Finally, we conducted numerical calculations to validate the effectiveness of the SSM. The effects of the electromechanical biasing fields and the tube geometry on the axisymmetric vibration characteristics were discussed in detail. From the numerical results, we obtained the following important conclusions:

1) The proposed SSM is a highly accurate and efficient method for studying the axisymmetric vibrations of SEA tubes under inhomogeneous biasing fields.

2) The manipulation of axisymmetric vibration behaviors of the neo-Hookean SEA tubes is feasible by tuning the electromechanical biasing fields except for the lowest torsional mode with linear displacement distribution along the radial direction.

3) By varying the tube geometrical size, the resonant frequencies of different modes for the neo-Hookean SEA tubes could be readily adjusted, except for the breathing mode and the lowest torsional mode that are independent of the length-to-thickness ratio and the outer-to-inner radius ratio, respectively.

This work provides not only a novel method (SSM) to derive the frequency equations of three-dimensional free vibrations of SEA cylindrical structures, but also demonstrates the electrostatic tunability of resonant frequency of SEA tubes with various geometrical sizes. The present investigation clearly indicates that it is feasible to use biasing fields to tune the small-amplitude vibration behaviors of SEA tubes, which should be beneficial to the design of tunable resonant devices consisting of SEA tubes.

**Acknowledgements**

This work was supported by a Government of Ireland Postdoctoral Fellowship from the Irish Research Council (No. GOIPD/2019/65) and the National Natural Science Foundation of China (Nos. 11872329 and 11621062). Partial supports from the Fundamental Research Funds for the Central Universities, PR China (No. 2016XZZX001-05) and the Shenzhen Scientific and Technological Fund for R&D, PR China (No. JCYJ20170816172316775) are also acknowledged.

## Appendix A. Theoretical background

### A.1 The theory of nonlinear electroelasticity

Consider a soft deformable continuous electroelastic body subjected to a static finite deformation. We denote the undeformed stress-free reference configuration at time $t_0$ by $\mathcal{B}_r$, and by $\partial\mathcal{B}_r$ and $\mathbf{N}$ the boundary and the outward unit normal, respectively. Any material point $X$ in $\mathcal{B}_r$ is identified by its position vector $\mathbf{X}$. An application of the external stimuli deforms the body so that the material point $X$ occupies a new position $\mathbf{x} = \boldsymbol{\chi}(\mathbf{X}, t)$ at time $t$ in the deformed or current configuration $\mathcal{B}_t$, with the boundary and the outward unit normal denoted by $\partial\mathcal{B}_t$ and $\mathbf{n}_t$, respectively. Here, the vector function $\boldsymbol{\chi}$ with a sufficiently regular property is defined for all points in $\mathcal{B}_r$. The deformation gradient tensor is defined as $\mathbf{F} = \partial\mathbf{x}/\partial\mathbf{X} = \mathrm{Grad}\boldsymbol{\chi}$, where 'Grad' is the gradient operator with respect to $\mathcal{B}_r$. The local measure of the volume change is $J = \det\mathbf{F} = 1$ for an incompressible material. The left and right Cauchy-Green strain tensors $\mathbf{b} = \mathbf{F}\mathbf{F}^{\mathrm{T}}$ and $\mathbf{c} = \mathbf{F}^{\mathrm{T}}\mathbf{F}$ are used as the deformation measures, where the superscript $^{\mathrm{T}}$ signifies the usual transpose operator.

Under the quasi-electrostatic approximation and in the absence of mechanical body forces, free body charges and currents, the equation of motion, Gauss's law and Faraday's law may be written as

$$\mathrm{div}\boldsymbol{\tau} = \rho\mathbf{x}_{,tt}, \quad \mathrm{div}\mathbf{D} = 0, \quad \mathrm{curl}\mathbf{E} = 0, \tag{A.1}$$

respectively, where $\rho$ is the material mass density which remains unchanged during the motion, the subscript $t$ following a comma denotes the material time derivative, 'div' and 'curl' are the divergence and curl operators in $\mathcal{B}_t$, respectively, and $\boldsymbol{\tau}$, $\mathbf{D}$ and $\mathbf{E}$ represent the total Cauchy stress tensor including the contribution of the electric body forces, the electric displacement and electric field vectors in $\mathcal{B}_t$, respectively.

For an incompressible material, the nonlinear constitutive relations can be expressed as

$$\mathbf{T} = \frac{\partial\Omega}{\partial\mathbf{F}} - p\mathbf{F}^{-1}, \quad \boldsymbol{\mathcal{E}} = \frac{\partial\Omega}{\partial\boldsymbol{\mathcal{D}}}, \tag{A.2}$$

where $\Omega(\mathbf{F}, \boldsymbol{\mathcal{D}})$ is the total energy density function per unit reference volume, $\mathbf{T} = \mathbf{F}^{-1}\boldsymbol{\tau}$, $\boldsymbol{\mathcal{D}} = \mathbf{F}^{-1}\mathbf{D}$ and $\boldsymbol{\mathcal{E}} = \mathbf{F}^{-1}\mathbf{E}$ are the total nominal stress tensor, the Lagrangian electric displacement and electric field vectors, respectively; $p$ is a Lagrange multiplier related to the incompressibility constraint. Due to incompressibility $(I_3 \equiv \det\mathbf{c} = 1)$, the energy density function $\Omega(\mathbf{F},\mathcal{D})$ depends the following five invariants only,

$$I_1 = \mathrm{tr}\mathbf{c},\quad I_2 = \frac{1}{2}\left[(\mathrm{tr}\mathbf{c})^2 - \mathrm{tr}\left(\mathbf{c}^2\right)\right],\quad I_4 = \mathcal{D}\cdot\mathcal{D},\quad I_5 = \mathcal{D}\cdot(\mathbf{c}\mathcal{D}),\quad I_6 = \mathcal{D}\cdot\left(\mathbf{c}^2\mathcal{D}\right). \tag{A.3}$$

Thus, the total stress tensor $\boldsymbol{\tau}$ and the Eulerian electric field vector $\mathbf{E}$ can be derived from Eqs. (A.2) and (A.3) as

$$\begin{aligned}&\boldsymbol{\tau} = 2\Omega_1\mathbf{b} + 2\Omega_2\left(I_1\mathbf{b} - \mathbf{b}^2\right) - p\mathbf{I} + 2\Omega_5\mathbf{D}\otimes\mathbf{D} + 2\Omega_6(\mathbf{D}\otimes\mathbf{bD} + \mathbf{bD}\otimes\mathbf{D}),\\ &\mathbf{E} = 2\left(\Omega_4\mathbf{b}^{-1}\mathbf{D} + \Omega_5\mathbf{D} + \Omega_6\mathbf{bD}\right),\end{aligned} \tag{A.4}$$

where $\Omega_m = \partial\Omega/\partial I_m\ (m=1,2,4,5,6)$.

Taking no account of the electrical quantities in the surrounding vacuum, the mechanical and electric boundary conditions to be satisfied on $\partial\mathcal{B}_t$ may be written as

$$\boldsymbol{\tau}\mathbf{n} = \mathbf{t}_\mathrm{a},\quad \mathbf{E}\times\mathbf{n}_t = 0,\quad \mathbf{D}\cdot\mathbf{n}_t = -\sigma_\mathrm{f}, \tag{A.5}$$

where $\mathbf{t}_\mathrm{a}$ is the applied mechanical traction vector per unit area of $\partial\mathcal{B}_t$ and $\sigma_\mathrm{f}$ is the free surface charge density on $\partial\mathcal{B}_t$.

### A.2 The linearized incremental theory

Now an incremental time-dependent perturbation $\dot{\mathbf{x}} = (\mathbf{X},t)$ along with an infinitesimal incremental electric displacement $\dot{\mathcal{D}}_0$ is superimposed upon a finitely deformed configuration $\mathcal{B}_0$ (with the boundary $\partial\mathcal{B}_0$ and the outward unit normal vector $\mathbf{n}$). Here, the incremental quantities are signified by the superposed dot. According to the incremental field theory [40], the updated Lagrangian form of the incremental governing equations can be written as

$$\mathrm{div}\mathbf{T}_0 = \rho\mathbf{u}_{,tt},\quad \mathrm{curl}\dot{\mathcal{E}}_0 = 0,\quad \mathrm{div}\dot{\mathcal{D}}_0 = 0, \tag{A.6}$$

where $\mathbf{u}(\mathbf{x},t) = \dot{\mathbf{x}}(\mathbf{X},t)$ is the incremental displacement vector and $\dot{\mathcal{D}}_0$, $\dot{\mathcal{E}}_0$ and $\mathbf{T}_0$ are the ‘push-forward’ versions of the corresponding Lagrangian increments. The resulting push-forward variables are identified with a subscript 0. The linearized incremental constitutive equations for incompressible SEA materials are

$$\dot{\mathbf{T}}_0 = \mathcal{A}_0\mathbf{H} + \boldsymbol{\Gamma}_0\dot{\mathcal{D}}_0 + p\mathbf{H} - \dot{p}\mathbf{I},\quad \dot{\mathcal{E}}_0 = \boldsymbol{\Gamma}_0^\mathrm{T}\mathbf{H} + \mathcal{R}_0\dot{\mathcal{D}}_0, \tag{A.7}$$

where $\mathbf{H} = \mathrm{grad}\mathbf{u}$ is the incremental displacement gradient tensor, $\dot{p}$ is the incremental Lagrange multiplier, and $\mathcal{A}_0$, $\boldsymbol{\Gamma}_0$ and $\mathcal{R}_0$ are, respectively, the fourth-, third- and second-order tensors, which are referred to as instantaneous electroelastic moduli tensors. In component form, $\boldsymbol{\mathcal{A}}_0$, $\boldsymbol{\Gamma}_0$ and $\boldsymbol{\mathcal{R}}_0$ are given by

$$\begin{gathered}\mathcal{A}_{0piqj} = F_{p\alpha}F_{q\beta}\mathcal{A}_{\alpha i\beta j} = \mathcal{A}_{0qjpi}, \quad \Gamma_{0piq} = F_{p\alpha}F^{-1}_{\beta q}\Gamma_{\alpha i\beta} = \Gamma_{0ipq}, \\ \mathcal{R}_{0ij} = F^{-1}_{\alpha i}F^{-1}_{\beta j}\mathcal{R}_{\alpha\beta} = \mathcal{R}_{0ji},\end{gathered} \tag{A.8}$$

with the referential electroelastic moduli tensors $\boldsymbol{\mathcal{A}}$, $\boldsymbol{\Gamma}$ and $\boldsymbol{\mathcal{R}}$ associated with $\Omega(\mathbf{F},\boldsymbol{\mathcal{D}})$ defined as

$$\mathcal{A}_{\alpha i\beta j} = \frac{\partial^2\Omega}{\partial F_{i\alpha}\partial F_{j\beta}}, \quad \Gamma_{\alpha i\beta} = \frac{\partial^2\Omega}{\partial F_{i\alpha}\partial \mathcal{D}_{\beta}}, \quad \mathcal{R}_{\alpha\beta} = \frac{\partial^2\Omega}{\partial \mathcal{D}_{\alpha}\partial \mathcal{D}_{\beta}}. \tag{A.9}$$

Additionally, the incremental incompressibility condition can be written as

$$\mathrm{div}\mathbf{u} = \mathrm{tr}\mathbf{H} = 0. \tag{A.10}$$

The updated Lagrangian incremental forms of the mechanical and electric boundary conditions are

$$\dot{\mathbf{T}}_0^{\mathrm{T}}\mathbf{n} = \dot{\mathbf{t}}_0^{\mathrm{A}}, \quad \dot{\boldsymbol{\mathcal{E}}}_0 \times \mathbf{n} = 0, \quad \dot{\boldsymbol{\mathcal{D}}}_0 \cdot \mathbf{n} = -\sigma_{\mathrm{F}0}, \tag{A.11}$$

where the increments of electrical variables in the surrounding vacuum have been neglected, $\dot{\mathbf{t}}_0^{\mathrm{A}}$ and $-\sigma_{\mathrm{F}0}$ are the updated Lagrangian incremental mechanical traction vector per unit area of $\partial\mathcal{B}_0$ and the incremental surface charge density on $\partial\mathcal{B}_0$, respectively.

## Appendix B. Elements of the system matrix M

The four partitioned 4 × 4 sub-matrices $\mathbf{M}_{ij}\left(i, j=1,2\right)$ of the system matrix $\mathbf{M}$ in the state equation (24) are given by [46]

$$\mathbf{M}_{11}=\begin{bmatrix} -\frac{1}{r} & -\frac{1}{r}\frac{\partial}{\partial\theta} & -\frac{\partial}{\partial z} & 0 \\ -\frac{c_{69}}{c_{66}}\frac{1}{r}\frac{\partial}{\partial\theta} & \frac{c_{69}}{c_{66}}\frac{1}{r} & 0 & -\frac{e_{26}}{c_{66}}\frac{1}{r}\frac{\partial}{\partial\theta} \\ -\frac{c_{58}}{c_{55}}\frac{\partial}{\partial z} & 0 & 0 & -\frac{e_{35}}{c_{55}}\frac{\partial}{\partial z} \\ \frac{q_1}{r} & \frac{q_1}{r}\frac{\partial}{\partial\theta} & q_2\frac{\partial}{\partial z} & 0 \end{bmatrix},\quad \mathbf{M}_{12}=\begin{bmatrix} 0 & 0 & 0 & 0 \\ 0 & \frac{1}{c_{66}} & 0 & 0 \\ 0 & 0 & \frac{1}{c_{55}} & 0 \\ 0 & 0 & 0 & -\frac{1}{\varepsilon_{11}} \end{bmatrix},$$

$$\mathbf{M}_{21}=\begin{bmatrix} \frac{\rho\partial^2}{\partial t^2}-\frac{q_7}{r^2}\frac{\partial^2}{\partial\theta^2}+\frac{q_3}{r^2}-q_9\frac{\partial^2}{\partial z^2} & \frac{q_3+q_7}{r^2}\frac{\partial}{\partial\theta} & \frac{q_4}{r}\frac{\partial}{\partial z} & -\left(\frac{q_8}{r^2}\frac{\partial^2}{\partial\theta^2}+q_{10}\frac{\partial^2}{\partial z^2}\right) \\ -\frac{q_3+q_7}{r^2}\frac{\partial}{\partial\theta} & \frac{\rho\partial^2}{\partial t^2}-\frac{q_3}{r^2}\frac{\partial^2}{\partial\theta^2}+\frac{q_7}{r^2}-c_{77}\frac{\partial^2}{\partial z^2} & -\frac{q_4+c_{47}}{r}\frac{\partial^2}{\partial\theta\partial z} & -\frac{q_8}{r^2}\frac{\partial}{\partial\theta} \\ -\frac{q_5}{r}\frac{\partial}{\partial z} & -\frac{c_{47}+q_5}{r}\frac{\partial^2}{\partial\theta\partial z} & \frac{\rho\partial^2}{\partial t^2}-q_6\frac{\partial^2}{\partial z^2}-\frac{c_{44}}{r^2}\frac{\partial^2}{\partial\theta^2} & 0 \\ -\left(\frac{q_8}{r^2}\frac{\partial^2}{\partial\theta^2}+q_{10}\frac{\partial^2}{\partial z^2}\right) & \frac{q_8}{r^2}\frac{\partial}{\partial\theta} & 0 & \frac{q_{11}}{r^2}\frac{\partial^2}{\partial\theta^2}+q_{12}\frac{\partial^2}{\partial z^2} \end{bmatrix},$$

$$\mathbf{M}_{22}=\begin{bmatrix} 0 & -\frac{c_{69}}{c_{66}}\frac{1}{r}\frac{\partial}{\partial\theta} & -\frac{c_{58}}{c_{55}}\frac{\partial}{\partial z} & -\frac{q_1}{r} \\ -\frac{1}{r}\frac{\partial}{\partial\theta} & -\left(\frac{c_{69}}{c_{66}}+1\right)\frac{1}{r} & 0 & \frac{q_1}{r}\frac{\partial}{\partial\theta} \\ -\frac{\partial}{\partial z} & 0 & -\frac{1}{r} & q_2\frac{\partial}{\partial z} \\ 0 & -\frac{e_{26}}{c_{66}}\frac{1}{r}\frac{\partial}{\partial\theta} & -\frac{e_{35}}{c_{55}}\frac{\partial}{\partial z} & -\frac{1}{r} \end{bmatrix}$$

where

$$q_1=\left(e_{12}-e_{11}\right)/\varepsilon_{11},\ q_2=\left(e_{13}-e_{11}\right)/\varepsilon_{11},\ n_1=c_{12}-c_{11}+e_{11}q_1,\ n_2=c_{13}-c_{11}+e_{11}q_2,$$
$$q_3=c_{22}-c_{12}+e_{12}q_1-n_1,\ q_4=c_{23}-c_{12}+e_{12}q_2-n_2,\ q_5=c_{23}-c_{13}+e_{13}q_1-n_1,$$
$$q_6=c_{33}-c_{13}+e_{13}q_2-n_2,\ q_7=c_{99}-c_{69}^2/c_{66}, q_8=e_{26}\left(1-c_{69}/c_{66}\right),\ q_9=c_{88}-c_{58}^2/c_{55},$$
$$q_{10}=e_{35}\left(1-c_{58}/c_{55}\right),\ q_{11}=e_{26}^2/c_{66}+\varepsilon_{22},\ q_{12}=e_{35}^2/c_{55}+\varepsilon_{33}.$$

**Appendix C. Frequency equations of non-axisymmetric vibrations in a pre-stretched hyperelastic tube**

In this appendix, we derive the frequency equations of non-axisymmetric vibrations in a pre-stretched hyperelastic tube by means of the *conventional displacement method*. Without the electromechanical coupling, the deformation in a hyperelastic tube is homogeneous with the relations, $\lambda_r = \lambda_\theta = \lambda_a = \lambda_b = \lambda_z^{-1/2}$ . The three-dimensional incremental governing equations for the pre-stretched hyperelastic tube can be obtained from Eqs. (18)-(21) by neglecting the electromechanical coupling terms. In fact, the governing equations for the hyperelastic tube can be deduced from those in Su et al. [47] for an SEA hollow cylinder with homogeneous biasing fields through a proper degenerate analysis.

Based on the basic governing equations without the electromechanical coupling obtained by Su et al. [47] (see their Eq. (41)), three displacement functions $\psi$, $G$ and $W$ are introduced to express the displacement components as

$$u_r = \frac{1}{r}\frac{\partial \psi}{\partial \theta} - \frac{\partial G}{\partial r}, \quad u_\theta = -\frac{\partial \psi}{\partial r} - \frac{1}{r}\frac{\partial G}{\partial \theta}, \quad u_z = W, \tag{C.1}$$

which, when combined with the relation $c_{11} - c_{12} - c_{69} = c_{66}$, yields

$$\begin{gathered}
\left(c_{66}\nabla^2 + c_{77}\frac{\partial^2}{\partial z^2} - \rho\frac{\partial^2}{\partial t^2}\right)\psi = 0, \quad -\nabla^2 G + \frac{\partial W}{\partial z} = 0, \\
\left[\left(c_{11} - c_{13} - c_{58}\right)\nabla^2 + c_{77}\frac{\partial^2}{\partial z^2} - \rho\frac{\partial^2}{\partial t^2}\right]G + \dot{p} = 0, \\
\left[c_{55}\nabla^2 + \left(c_{33} - c_{13} - c_{58}\right)\frac{\partial^2}{\partial z^2} - \rho\frac{\partial^2}{\partial t^2}\right]W - \frac{\partial \dot{p}}{\partial z} = 0,
\end{gathered} \tag{C.2}$$

where $\nabla^2 = \partial^2/\partial r^2 + (1/r)\partial/\partial r + (1/r^2)\partial^2/\partial\theta^2$ is the two-dimensional Laplace operator.

We look for the vibration solutions to Eq. (C.2) in the form:

$$\begin{aligned}
&\psi = \bar{\psi}(r)\sin(m\theta)\cos(n\pi\zeta)e^{\mathrm{i}\omega t}, \quad G = \bar{G}(r)\cos(m\theta)\cos(n\pi\zeta)e^{\mathrm{i}\omega t}, \\
&W = \bar{W}(r)\cos(m\theta)\sin(n\pi\zeta)e^{\mathrm{i}\omega t}, \quad \dot{p} = \bar{p}(r)\cos(m\theta)\cos(n\pi\zeta)e^{\mathrm{i}\omega t},
\end{aligned} \tag{C.3}$$

where $\zeta = z/l$ is the dimensionless axial coordinate and $m$ is the circumferential wave number. Note that Eq. (C.3), while satisfying the generalized rigidly supported conditions (28) at the tube ends, represents the non-axisymmetric vibrations and can be reduced to the axisymmetric vibrations by setting $m = 0$. Substituting Eq. (C.3) into Eq. (C.2), we obtain

$$\left(\Lambda+\alpha_3^2\right)\bar{\psi}=0,\quad -\Lambda\bar{G}+\gamma\bar{W}=0,$$
$$\left[\left(c_{11}-c_{13}-c_{58}\right)\Lambda+\rho\omega^2-\gamma^2 c_{77}\right]\bar{G}+\bar{p}=0, \tag{C.4}$$
$$\left[c_{55}\Lambda+\rho\omega^2-\left(c_{33}-c_{13}-c_{58}\right)\gamma^2\right]\bar{W}+\gamma\bar{p}=0,$$

where $\Lambda=\mathrm{d}^2/\mathrm{d}r^2+(1/r)\mathrm{d}/\mathrm{d}r-m^2/r^2$, $\alpha_3^2=(\rho\omega^2-\gamma^2 c_{77})/c_{66}$ and $\gamma=n\pi/l$.

It is apparent that Eq. (C.4)$_1$ is a Bessel equation of order $m$, and its solution is

$$\bar{\psi}=A_3 J_m\left(\alpha_3 r\right)+B_3 Y_m\left(\alpha_3 r\right), \tag{C.5}$$

where $J_m(\cdot)$ and $Y_m(\cdot)$ are the Bessel functions of the first and second kinds of order $m$, respectively, and $A_3$ and $B_3$ are arbitrary constants to be determined from the boundary conditions. As for the remaining equations in Eq. (C.4), their solution can be assumed as [47]

$$\begin{Bmatrix}\bar{G}\\ \bar{W}\\ \bar{p}\end{Bmatrix}=J_m\left(\alpha r\right)\begin{Bmatrix}C_1\\ C_2\\ C_3\end{Bmatrix}+Y_m\left(\alpha r\right)\begin{Bmatrix}D_1\\ D_2\\ D_3\end{Bmatrix}, \tag{C.6}$$

where $\alpha$ is the radial wave number related to the other three functions $\bar{G}$, $\bar{W}$ and $\bar{p}$, and $C_j$ and $D_j$ $\left(j=1-3\right)$ are undetermined constants. Inserting Eq. (C.6) into Eq. (C.4)$_{2\text{-}4}$ and ensuring the non-trivial solutions, the determinant of the coefficient matrix associated with $C_j$ and $D_j$ must be zero, which results in the following characteristic equation:

$$\begin{vmatrix}g_{11} & 0 & 1\\ 0 & g_{22} & g_{23}\\ g_{31} & g_{32} & 0\end{vmatrix}=0, \tag{C.7}$$

where

$$g_{11}=\rho\omega^2-c_{77}\gamma^2-\alpha^2\left(c_{11}-c_{13}-c_{58}\right),$$
$$g_{22}=\rho\omega^2-\left(c_{33}-c_{13}-c_{58}\right)\gamma^2-\alpha^2 c_{55}, \tag{C.8}$$
$$g_{23}=g_{32}=\gamma,\quad g_{31}=\alpha^2.$$

For the prescribed $n$ and $\omega$, the characteristic equation can yield two different values of $\alpha$ with $\mathrm{Re}[\alpha_j]>0$ or $\mathrm{Re}[\alpha_j]=0$ and $\mathrm{Im}[\alpha_j]>0$. The complete vibration solutions can be expressed as

$$\begin{Bmatrix}\bar{G}\\ \bar{W}\\ \bar{p}\end{Bmatrix}=\sum_{j=1}^{2}\begin{Bmatrix}1\\ \beta_{1j}\\ \beta_{2j}\end{Bmatrix}\left[A_j J_m\left(\alpha_j r\right)+B_j Y_m\left(\alpha_j r\right)\right], \tag{C.9}$$

where $A_j$ and $B_j$ $(j=1-2)$ are undetermined constants and the ratio $\beta_{1j}$ and $\beta_{2j}$ between different constants $C_{ij}$ or $D_{ij}$ $(i=1-3)$ are obtained as

$$\beta_{1j}=-\alpha_j^2/\gamma,\quad \beta_{2j}=-\left[\rho\omega^2-\left(c_{33}-c_{13}-c_{58}\right)\gamma^2-\alpha_j^2 c_{55}\right]\beta_{1j}/\gamma,\quad \left(j=1-2\right).\quad \text{(C.10)}$$

After substituting Eqs. (C.5) and (C.9) into Eqs. (C.1) and $(21)_{1,4,8}$, we obtain the incremental displacement and transverse stress components. Then, according to the mechanical part of incremental boundary conditions (38) and for the non-trivial solutions to exist, we can obtain the frequency equation of the non-axisymmetric vibrations as

$$\left|d_{ij}\right|=0,\quad \left(i,j=1-6\right),\quad \text{(C.11)}$$

where the elements $d_{ij}$ $(i=1-3)$ of the first three rows of the determinant corresponding to the boundary conditions on the outer surface $r=b$ are written as

$$\begin{aligned}
d_{1j}&=-c_{11}Z_m''\left(\alpha_j b\right)+c_{12}\frac{1}{b}\left[\frac{m^2}{b}Z_m\left(\alpha_j b\right)-Z_m'\left(\alpha_j b\right)\right]+\left(c_{13}\gamma\beta_{1j}-\beta_{2j}\right)Z_m\left(\alpha_j b\right),\\
d_{13}&=c_{11}\left[\frac{m}{b}J_m'\left(\alpha_3 b\right)-\frac{m}{b^2}J_m\left(\alpha_3 b\right)\right]+c_{12}\frac{1}{b}\left[\frac{m}{b}J_m\left(\alpha_3 b\right)-mJ_m'\left(\alpha_3 b\right)\right],\\
d_{16}&=c_{11}\left[\frac{m}{b}Y_m'\left(\alpha_3 b\right)-\frac{m}{b^2}Y_m\left(\alpha_3 b\right)\right]+c_{12}\frac{1}{b}\left[\frac{m}{b}Y_m\left(\alpha_3 b\right)-mY_m'\left(\alpha_3 b\right)\right],\\
d_{2j}&=\frac{c_{69}}{b}\left[mZ_m'\left(\alpha_j b\right)-\frac{m}{b}Z_m\left(\alpha_j b\right)\right]+c_{66}\left[\frac{m}{b}Z_m'\left(\alpha_j b\right)-\frac{m}{b^2}Z_m\left(\alpha_j b\right)\right],\qquad \text{(C.12)}\\
d_{23}&=\frac{c_{69}}{b}\left[-\frac{m^2}{b}J_m\left(\alpha_3 b\right)+J_m'\left(\alpha_3 b\right)\right]-c_{66}J_m''\left(\alpha_3 b\right),\\
d_{26}&=\frac{c_{69}}{b}\left[-\frac{m^2}{b}Y_m\left(\alpha_3 b\right)+Y_m'\left(\alpha_3 b\right)\right]-c_{66}Y_m''\left(\alpha_3 b\right),\\
d_{3j}&=c_{55}\beta_{1j}Z_m'\left(\alpha_j b\right)+c_{58}\gamma Z_m'\left(\alpha_j b\right),\ \ d_{33}=-c_{58}\gamma\frac{m}{b}J_m\left(\alpha_3 b\right),\ \ d_{36}=-c_{58}\gamma\frac{m}{b}Y_m\left(\alpha_3 b\right),
\end{aligned}$$

where $j=1,2,4,5$, the prime denotes differentiation with respect to $r$, $Z(\cdot)=J(\cdot)$ for $j=1,2$ and $Z(\cdot)=Y(\cdot)$ for $j=4,5$. In addition, the notations $\alpha_{j+3}=\alpha_j$ and $\beta_{i(j+3)}=\beta_{ij}$ $(i,j=1,2)$ have been adopted in Eq. (C.12). For the boundary conditions on the inner surface $r=a$, we can use the inner radius $a$ to replace the outer radius $b$ in Eq. (C.12) to obtain the elements $d_{ij}$ $(i=4-6)$ of the last three rows of the determinant (C.11).

For the axisymmetric vibrations with $m=0$, the frequency equation (C.11) can be decomposed as

$$\left|d_{ij}\right| = \begin{vmatrix} d_{11} & d_{12} & d_{14} & d_{15} \\ d_{31} & d_{32} & d_{34} & d_{35} \\ d_{41} & d_{42} & d_{44} & d_{45} \\ d_{61} & d_{62} & d_{64} & d_{65} \end{vmatrix} \begin{vmatrix} d_{23} & d_{26} \\ d_{53} & d_{56} \end{vmatrix} = S_1 \cdot S_2 = 0, \tag{C.13}$$

where $S_1 = 0$ and $S_2 = 0$ represent the axisymmetric longitudinal vibration (L vibration) and the purely torsional vibration (T vibration), respectively, of the pre-stretched hyperelastic tube.

Now consider neo-Hookean hyperelastic materials whose strain-energy function is characterized by Eq. (13) with $I_5 = 0$. We can obtain the necessary instantaneous elastic moduli and effective material parameters from Appendix B in Wu et al. [46] and Eq. (23) as

$$\begin{aligned} &\mathcal{A}_{01111} = \mathcal{A}_{01212} = \mu\lambda_z^{-1}, \quad \mathcal{A}_{01221} = \mathcal{A}_{01122} = 0, \quad p = \mu\lambda_\theta^2 = \mu\lambda_z^{-1}, \\ &c_{11} = 2\mu\lambda_z^{-1}, \quad c_{12} = 0, \quad c_{66} = c_{69} = \mu\lambda_z^{-1}, \quad c_{77} = \mu\lambda_z^2, \end{aligned} \tag{C.14}$$

which yields $\bar{\alpha}_3^2 = \alpha_3^2 H^2 = \lambda_z\left(\varpi^2 - \kappa^2\right)$ with $\varpi = \omega H / \sqrt{\mu / \rho}$ and $\kappa = n\pi H / L$. Substituting Eq. (C.14) into Eq. (C.13), we rewrite the elements associated with the purely torsional vibration, in the dimensionless form, as

$$\begin{aligned} \bar{d}_{23} &= d_{23}\frac{H^2}{\mu} = \bar{\alpha}_{3\lambda}^2 J_0\left(\frac{\bar{\alpha}_{3\lambda}}{1-\eta}\right) - 2(1-\eta)\bar{\alpha}_{3\lambda} J_1\left(\frac{\bar{\alpha}_{3\lambda}}{1-\eta}\right), \\ \bar{d}_{26} &= d_{26}\frac{H^2}{\mu} = \bar{\alpha}_{3\lambda}^2 Y_0\left(\frac{\bar{\alpha}_{3\lambda}}{1-\eta}\right) - 2(1-\eta)\bar{\alpha}_{3\lambda} Y_1\left(\frac{\bar{\alpha}_{3\lambda}}{1-\eta}\right), \\ \bar{d}_{53} &= d_{53}\frac{H^2}{\mu} = \bar{\alpha}_{3\lambda}^2 J_0\left(\frac{\eta\bar{\alpha}_{3\lambda}}{1-\eta}\right) - 2\left(\eta^{-1}-1\right)\bar{\alpha}_{3\lambda} J_1\left(\frac{\eta\bar{\alpha}_{3\lambda}}{1-\eta}\right), \\ \bar{d}_{56} &= d_{56}\frac{H^2}{\mu} = \bar{\alpha}_{3\lambda}^2 Y_0\left(\frac{\eta\bar{\alpha}_{3\lambda}}{1-\eta}\right) - 2\left(\eta^{-1}-1\right)\bar{\alpha}_{3\lambda} Y_1\left(\frac{\eta\bar{\alpha}_{3\lambda}}{1-\eta}\right), \end{aligned} \tag{C.15}$$

where $\bar{\alpha}_{3\lambda} = \bar{\alpha}_3\lambda_1 = \bar{\alpha}_3\lambda_z^{-1/2} = \sqrt{\varpi^2 - \kappa^2}$. It is obvious from Eq. (C.15) that the resonant frequency of the purely torsional vibration is independent of the axial pre-stretch $\lambda_z$, but depends on the length-to-thickness ratio $L/H$ in $\kappa$ and the inner-to-outer radius ratio $\eta = A/B$. After some manipulations, the frequency equation $S_2 = 0$ becomes

$$\bar{\alpha}_{3\lambda}^4\left[J_2\left(\frac{\bar{\alpha}_{3\lambda}}{1-\eta}\right)Y_2\left(\frac{\eta\bar{\alpha}_{3\lambda}}{1-\eta}\right) - J_2\left(\frac{\eta\bar{\alpha}_{3\lambda}}{1-\eta}\right)Y_2\left(\frac{\bar{\alpha}_{3\lambda}}{1-\eta}\right)\right] = 0. \tag{C.16}$$

Thus, we note that $\bar{\alpha}_{3\lambda}^2 = \varpi^2 - \kappa^2 = 0$ is one of the solutions to the frequency equation of the purely torsional vibration, which is only determined by $L/H$. In fact, the torsional displacement corresponding to $\varpi = \kappa$ is proportional to the radius, and thus the vibration is a rotation of each cross-section of the tube as a whole about its center, which is similar to the torsional waves in an isotropic elastic cylinder [54].

For the breathing mode with $m = u_\theta = u_z = 0$, we have from Eqs. (C.1) and (C.3)

$$u_r = \bar{u}(r)e^{\mathrm{i}\omega t}, \quad u_\theta = 0, \quad u_z = 0, \quad \dot{p} = \bar{p}(r)e^{\mathrm{i}\omega t}. \tag{C.17}$$

Substituting it into the incremental incompressibility condition (20) and the incremental governing equations (18) gives

$$\begin{gathered} \bar{u}'(r) = -\frac{1}{r}\bar{u}(r), \quad \bar{u}''(r) = \frac{2}{r^2}\bar{u}(r), \\ c_{11}\left[\bar{u}''(r) + \frac{\bar{u}'(r)}{r} - \frac{\bar{u}(r)}{r^2}\right] - \bar{p}'(r) = -\bar{u}(r)\rho\omega^2. \end{gathered} \tag{C.18}$$

Thus, the solutions to Eq. (C.18) can be written as

$$\bar{u} = A_4 / r, \quad \bar{p} = A_4\rho\omega^2 \ln r + B_4, \tag{C.19}$$

where $A_4$ and $B_4$ are the undetermined constants. Utilizing the incremental boundary conditions $\dot{T}_{0rr}\big|_{r=a,b} = (c_{12}u_r / r + c_{11}u_r' - \dot{p})\big|_{r=a,b} = 0$, we can obtain the frequency equation for the breathing mode as

$$\rho\omega^2 = \left(c_{12} - c_{11}\right)\left(\frac{1}{a^2} - \frac{1}{b^2}\right) / \ln\frac{a}{b}, \tag{C.20}$$

which, when combined with Eq. (C.14) and $\bar{\eta} = \eta$, leads to the frequency equation for the pre-stretched neo-Hookean hyperelastic tube as

$$\varpi^2 = -2(1-\eta)^2\left(1-\eta^2\right) / \left(\eta^2 \ln\eta\right). \tag{C.21}$$

**Appendix D. Frequency equation of the purely torsional mode with linear displacement distribution in an SEA tube**

For the purely torsional vibration with $m = u_r = u_z = \dot{p} = \dot{\varphi} = 0$, the only non-zero displacement component is

$$u_\theta = \bar{v}(r)\cos(n\pi\zeta)e^{\mathrm{i}\omega t}, \tag{D.1}$$

which satisfies the generalized rigidly supported conditions (28) at the tube ends. Therefore, the incremental incompressibility condition (20) is satisfied automatically. Substituting Eq. (D.1) into Eqs. (21)-(22), we can obtain the non-zero stress and electric displacement components as

$$\begin{aligned}&\dot{T}_{0r\theta} = c_{66}\frac{\partial u_\theta}{\partial r} - c_{69}\frac{u_\theta}{r}, \quad \dot{T}_{0\theta r} = c_{69}\frac{\partial u_\theta}{\partial r} - c_{99}\frac{u_\theta}{r},\\ &\dot{T}_{0\theta z} = c_{47}\frac{\partial u_\theta}{\partial z}, \quad \dot{T}_{0z\theta} = c_{77}\frac{\partial u_\theta}{\partial z}, \quad \dot{\mathcal{D}}_{0\theta} = e_{26}\left(\frac{\partial u_\theta}{\partial r} - \frac{u_\theta}{r}\right).\end{aligned} \tag{D.2}$$

Consequently, the incremental governing equations (17) and $(18)_{1,3}$ are also satisfied automatically. Inserting Eq. (D.2) into Eq. $(18)_2$ yields

$$c_{66}\bar{v}''(r) + \left(c'_{66} + \frac{c_{66}}{r}\right)\bar{v}'(r) - \left(c'_{69} + \frac{c_{99}}{r}\right)\frac{\bar{v}(r)}{r} + \left(\rho\omega^2 - c_{77}\gamma^2\right)\bar{v}(r) = 0, \tag{D.3}$$

where $\gamma = n\pi/l$ and the prime signifies differentiation with respect to $r$.

Note that based on the incremental form of the total Cauchy stress symmetry condition $\mathbf{FT} = (\mathbf{FT})^{\mathrm{T}}$, the connections between the components of $\mathcal{A}_0$ and $\boldsymbol{\tau}$ for an incompressible material may be obtained as [27, 46]

$$\mathcal{A}_{0jisk} - \mathcal{A}_{0ijsk} = \left(\tau_{js} + p\delta_{js}\right)\delta_{ik} - \left(\tau_{is} + p\delta_{is}\right)\delta_{jk}, \tag{D.4}$$

which, when combined with Eq. (23), provides

$$c_{66} - c_{69} = \tau_{11}, \quad c_{99} - c_{69} = \tau_{22}, \quad c_{99} = \tau_{22} - \tau_{11} + c_{66}. \tag{D.5}$$

Thus, using Eq. (D.5) and the equilibrium equation $(6)_2$ with $\tau_{rr} = \tau_{11}$ and $\tau_{\theta\theta} = \tau_{22}$, Eq. (D.3) can be rewritten as

$$c_{66}\bar{v}''(r) + \left(c'_{66} + \frac{c_{66}}{r}\right)\bar{v}'(r) - \left(c'_{66} + \frac{c_{66}}{r}\right)\frac{\bar{v}(r)}{r} + \alpha_3^2 c_{66}\bar{v}(r) = 0, \tag{D.6}$$

where $\alpha_3^2 = (\rho\omega^2 - \gamma^2 c_{77})/c_{66}$. It is obvious that the solution to Eq. (D.6) is difficult to obtain for $\alpha_3^2 \neq 0$. However, if $\alpha_3^2 = 0$, Eq. (D.6) becomes

$$c_{66}\left[\bar{v}''(r)+\frac{1}{r}\bar{v}'(r)-\frac{1}{r^2}\bar{v}(r)\right]=c_{66}'\left[\frac{\bar{v}(r)}{r}-\bar{v}'(r)\right]. \tag{D.7}$$

Now we assume that the left-hand side of Eq. (D.7) is equal to zero, i.e.,

$$\bar{v}''(r)+\frac{1}{r}\bar{v}'(r)-\frac{1}{r^2}\bar{v}(r)=0, \tag{D.8}$$

which is the classical Euler equation with the general solution $\bar{v}(r)=A_5r^{-1}+B_5r$, where $A_5$ and $B_5$ are the undetermined constants. In order to make the right-hand side of Eq. (D.7) vanish, we take $A_5=0$, which yields the solution of $\bar{v}(r)$ and $u_\theta$ as

$$\bar{v}(r)=B_5r,\quad u_\theta=B_5r\cos(n\pi\zeta)e^{\mathrm{i}\omega t}. \tag{D.9}$$

Thus, the solution (D.9) satisfies the governing equation (D.7).

Inserting Eq. (D.9) into Eq. (D.2)$_1$, we have $\dot{T}_{0r\theta}=B_5(c_{66}-c_{69})\cos(n\pi\zeta)e^{\mathrm{i}\omega t}$. Using Eq. (D.5)$_1$ and $\tau_{11}|_{r=a,b}=\tau_{rr}|_{r=a,b}=0$, the incremental boundary conditions $\dot{T}_{0r\theta}|_{r=a,b}=0$ are satisfied automatically. As a result, regardless of the inhomogeneous biasing fields, Eq. (D.9) is indeed a solution of the purely torsional vibration in the SEA tube with $\alpha_3^2=0$ determining its frequency equation. Specifically, the torsional mode related to $\alpha_3^2=0$ exhibits linear displacement distribution along the radial direction.

Particularly, for the neo-Hookean ideal dielectric model (13), the frequency equation $\alpha_3^2=0$ can be rewritten as

$$\varpi^2\equiv\rho\omega^2H^2/\mu=(n\pi H/L)^2\equiv\kappa^2, \tag{D.10}$$

which is independent of the axial pre-stretch $\lambda_z$ and the inner-to-outer radius ratio $\eta=A/B$, and depends only on the length-to-thickness ratio $L/H$. This phenomenon is analogous to the hyperelastic case described in Appendix C.